\documentclass[twocolumn]{aastex62}

\usepackage{xspace}
\usepackage{graphicx}
\usepackage{natbib}
\usepackage{amsmath}
\usepackage{afterpage}
\usepackage{longtable}

\newcommand{\nraoblurb}{The National Radio Astronomy Observatory is a facility of the National Science Foundation operated under cooperative agreement by Associated Universities, Inc.}
\newcommand{\gboblurb}{The Green Bank Observatory is a facility of the National Science Foundation operated under cooperative agreement by Associated Universities, Inc.}

\newcommand{\hide}[1]{}

\newcommand{\kms}{\ensuremath{\,{\rm km\,s^{-1}}}\xspace}

\newcommand{\mhz}{\ensuremath{\,{\rm MHz}}\xspace}

\newcommand{\degree}{\ensuremath{^\circ}\xspace}

\newcommand{\lsun}{\ensuremath{\,L_\odot}\xspace}     
 
\newcommand{\hii}{{\rm H\,{\footnotesize II}}\xspace}

\newcommand{\heh}{\ensuremath{N(\textnormal{He}^+)/N(\textnormal{H}^+)}\xspace}

\shorttitle{GDIGS: Tracing the DIG around W43}
\shortauthors{Luisi et al.}

\begin{document}

\title{The GBT Diffuse Ionized Gas Survey: Tracing the Diffuse Ionized Gas around the Giant H{\footnotesize II} Region W43}

\author[0000-0001-8061-216X]{Matteo~Luisi}
\affiliation{Department of Physics and Astronomy, West Virginia University, Morgantown WV 26506, USA}
\affiliation{Center for Gravitational Waves and Cosmology, West Virginia University, Chestnut Ridge Research Building, Morgantown WV 26505, USA}

\author[0000-0001-8800-1793]{L.~D.~Anderson}
\affiliation{Department of Physics and Astronomy, West Virginia University, Morgantown WV 26506, USA}
\affiliation{Center for Gravitational Waves and Cosmology, West Virginia University, Chestnut Ridge Research Building, Morgantown WV 26505, USA}
\affiliation{Adjunct Astronomer at the Green Bank Observatory, P.O. Box 2, Green Bank WV 24944, USA}

\author[0000-0002-1311-8839]{Bin~Liu}
\affiliation{CAS Key Laboratory of FAST, National Astronomical Observatories, Chinese Academy of Sciences, Beijing 100101, People's Republic of China}

\author[0000-0002-2465-7803]{Dana~S.~Balser}
\affiliation{National Radio Astronomy Observatory, 520 Edgemont Road, Charlottesville VA 22903-2475, USA}

\author[0000-0003-4866-460X]{T.~M.~Bania}
\affiliation{Institute for Astrophysical Research, Department of Astronomy, Boston University, 725 Commonwealth Ave., Boston MA 02215, USA}

\author[0000-0003-0640-7787]{Trey~V.~Wenger}
\affiliation{Dominion Radio Astrophysical Observatory, Herzberg Astronomy and Astrophysics Research Centre, National Research Council Canada, P.O.~Box 248, Penticton, BC V2A 6J9, Canada}

\author[0000-0002-9947-6396]{L.~M.~Haffner}
\affiliation{Department of Physical Sciences, Embry-Riddle Aeronautical University, Daytona Beach FL 32114, USA}

\begin{abstract}
The Green Bank Telescope (GBT) Diffuse Ionized Gas Survey (GDIGS) is a fully-sampled radio recombination line (RRL) survey of the inner Galaxy at C-band ($4-8$\,GHz). We average together $\sim 15$ Hn$\alpha$ RRLs within the receiver bandpass to improve the spectral signal-to-noise ratio. The average beam size for the RRL observations at these frequencies is $\sim 2$\arcmin. We grid these data to have spatial and velocity spacings of 30\arcsec\ and 0.5\kms, respectively. Here we discuss the first RRL data from GDIGS: a six square-degree-area surrounding the Galactic \hii\ region complex W43. We attempt to create a map devoid of emission from discrete \hii\ regions and detect RRL emission from the diffuse ionized gas (DIG) across nearly the entire mapped area. We estimate the intensity of the DIG emission by a simple empirical model, taking only the \hii\ region locations, angular sizes, and RRL intensities into account. The DIG emission is predominantly found at two distinct velocities: $\sim 40\kms$ and $\sim 100\kms$. While the 100\kms component is associated with W43 at a distance of $\sim 6$\,kpc, the origin of the 40\kms component is less clear. Since the distribution of the 40\kms emission cannot be adequately explained by ionizing sources at the same velocity, we hypothesize that the plasma at the two velocity components is interacting, placing the 40\kms DIG at a similar distance as the 100\kms emission. We find a correlation between dust temperature and integrated RRL intensity, suggesting that the same radiation field that heats the dust also maintains the ionization of the DIG.
\end{abstract}


\section{Introduction}\label{sec:intro}

The warm ionized medium (WIM), is a low-density ($\sim$0.1\,cm$^{-3}$), diffuse ionized gas component of the interstellar medium (ISM), first proposed by \citet{Hoyle1963}. It makes up $\sim$20\% of the total Milky Way gas mass and $>$90\% of its ionized gas mass \citep{Reynolds1991}. With temperatures ranging from 6000 to 10,000\,K \citep{Haffner2009} and a hydrogen ionization ratio $n(\textnormal{H}^+)/n(\textnormal{H}^0) \geq 13$ at the upper end of this temperature range \citep{Reynolds1998}, it provides a major source of pressure at the Galactic midplane \citep{Boulares1990}. The ``Diffuse Ionized Gas" (DIG) is often quoted in the same context as the WIM, with the WIM sometimes referring to the lower-density gas that cannot be traced to individual source regions.

Most WIM/DIG studies to date have been conducted by observing H$\alpha$ emission. Although H$\alpha$ is very bright compared to other ionized gas tracers, it suffers from extinction, limiting the distance to which the DIG in the inner Galaxy ($90\degree > \ell > -90\degree$) can be studied. Photometric H$\alpha$ surveys of the DIG include the Southern H-Alpha Sky Survey Atlas \citep[SHASSA;][]{Gaustad2001} and the Virginia Tech Spectral-Line Survey \citep[VTSS;][]{Dennison1998}. With its coarse spatial and spectral resolution, the all-sky Wisconsin H-Alpha Mapper \citep[WHAM;][]{Reynolds1998} provides a clear view of the global properties of the DIG, but is inappropriate for studying the inner Galaxy mid-plane where the majority of massive stars are located. Using the Near Infrared Camera and Multi-Object Spectrometer (NICMOS) instrument aboard the Hubble Space Telescope, \citet{Wang2010} report on a high-resolution photometric Paschen-$\alpha$ survey of the Galactic Center. Due to the longer wavelength, their data suffer less from extinction than H$\alpha$ observations. The lack of velocity information, however, provides a challenge in disentangling DIG emission along the line of sight.

Radio observations give us an opportunity to investigate the distribution of the DIG throughout the Galactic disk. Despite its low density, $\sim$80--90\% of the total Galactic free-free emission is believed to come from the DIG and thus the DIG is a major source of radio continuum and radio recombination line (RRL) emission \citep{Reynolds1984}. There have been numerous studies of the DIG using low frequency radio observations \citep[e.g.,][]{Roshi2000}, with a spatial resolution of $\sim 2\degree$. \citet{Liu2013} reports on an ongoing beam-sampled 1.4 GHz RRL survey using the 305-m Arecibo Radio Telescope called ``SIGGMA." These data extend from 65\degree $> \ell > 32$\degree, $|b| < 1.5$\degree. The spatial resolution is $\sim 3\arcmin$, and the spectral resolution is $\sim 5$\kms. Because of the survey strategy, some large-scale DIG emission is filtered out in SIGGMA. The other extant RRL survey with sub-degree resolution is the 1.4\,GHz survey of \citet{Alves2015}. Due to limited sensitivity, however, the data were binned to 20\kms spectral resolution at $15\arcmin$ spatial resolution.

While the most likely source of ionizing photons is high-mass stars \citep[e.g.,][]{Domgoergen1994}, the processes of how the DIG maintains its ionization are still not fully understood. For example, the DIG is found to be in a lower ionization state compared to the plasma in \hii\ regions \citep{Madsen2006}. Furthermore, the radiation field emitted by O-stars changes as the radiation transitions from the \hii\ region through the photodissociation region \citep[PDR; e.g.,][]{Hoopes2003}.  Using Monte Carlo photoionization simulations, \citet{Wood2004} argue that helium-ionizing photons in density bounded \hii\ regions are suppressed as radiation escapes from the region, while the hydrogen-ionizing continuum hardens. The suppression of helium-ionizing photons should lead to a decrease in the \heh ionic abundance ratio with distance from the region, an effect observed for a sample of Galactic \hii\ regions \citep{Luisi2016,Luisi2019}. It is not clear, however, where this suppression takes place, and what mechanisms are responsible.

It further remains unclear how the radiation from O-stars in \hii\ regions is able to propagate across the kiloparsec size-scales required given the observed DIG distribution. The radiation must first escape through the \hii\ region PDRs, and \hii\ regions themselves should be surrounded by ``halos" of diffuse ionized gas. These halos have been observed around individual Galactic \hii\ regions \citep{Deharveng2009,Anderson2015}, but not in a systematic study. Although the halos are sometimes thought of as being separate from the DIG \citep[e.g.,][]{McKee1997}, here we do not draw a distinction. 

\hii\ regions are known to leak a significant fraction of their ionizing photons into the ISM, but the exact percentage is only known very roughly. \citet{Oey1997} find that up to 50\% of the ionizing radiation escapes Large Magellanic Cloud (LMC) nebulae. \citet[][2002]{Zurita2000} simulate escape fractions from \hii\ regions in the galaxy NGC\,157 to match the observed H$\alpha$ distribution, and find good agreement when 30\% of Lyman continuum photons escape from each \hii\ region and propagate through the ISM isotropically. These studies have so far only focused on the most luminous regions. Recent investigations of less luminous \hii\ regions find a lower percentage of leaking photons than that of previous works on more luminous regions: 15$\pm 5\%$ for NGC\,7538 \citep{Luisi2016} and 25$\pm 10\%$ for RCW\,120 \citep{Anderson2015}. These results may indicate that giant \hii\ regions have a much larger effect in maintaining the DIG than compact \hii\ regions and may be the primary source of ionizing photons in the ISM, as previously suggested by, e.g., \citet{Zurita2000}.

Distinguishing the DIG from \hii\ region emission is a major challenge in analyzing RRL emission from the Galactic plane. In the \hii\ Region Discovery Survey (HRDS), \citet{Bania2010} observed discrete \hii\ regions and find that nearly 30\% of all targets have two or more separate RRL components at different velocities. Using subsequent observations, \citet{Anderson2015a} show that one of these components is from the DIG and the other is from the compact \hii\ region that was targeted. In a follow-up study, \citet{Luisi2017} analyze the DIG-only components from the HRDS from $40\degree > \ell > 18\degree$ and $|b| < 1\degree$ and find that the intensity of diffuse Spitzer GLIMPSE 8.0\,$\mu$m emission is correlated with both the locations of discrete \hii\ regions and the intensity of the diffuse RRL emission. This suggests that the soft UV photons responsible for creating the infrared (IR) emission through excitation of polycyclic aromatic hydrocarbons (PAHs) may have a similar origin as the more energetic photons causing the RRL emission.

The Green Bank Telescope (GBT) Diffuse Ionized Gas Survey (GDIGS) is an ongoing RRL survey of the Milky Way disk that is investigating the distribution and properties of the DIG. GDIGS will encompass $\sim 46$\,deg$^2$ from $32.3\degree > \ell > -5\degree$ and $|b| < 0.5\degree$ once finished. In addition, GDIGS will make several explorations above and below the Galactic plane around active star formation regions. The observing frequencies of 4--8\,GHz make GDIGS essentially extinction-free, with higher spatial and spectral resolution compared to previous large-scale RRL surveys. 

Here we show and analyze our GDIGS data of the \hii\ region complex W43, one of the most active zones of star formation in the inner Galaxy \citep[e.g.,][]{NguyenLuong2011}. W43 is located at a parallax distance of $5.49^{+0.39}_{-0.34}$\,kpc \citep{Zhang2014}, close to the near end of the Galactic bar and the inner Scutum arm, a region of complex gas dynamics \citep[e.g.,][]{Benjamin2005,Wegg2015}. In fact, \citet{NguyenLuong2011} suggest that the W43 complex may have formed due to its particular location via converging gas flows. In a GBT study of the inner-Galaxy DIG, \citet{Luisi2017} show that the diffuse gas near W43 is predominantly found at two distinct velocities: $\sim 40$\kms and $\sim 100$\kms. It is still uncertain whether these two velocity components originate from two distinct distances or whether they may be caused by noncircular streaming motions near the end of the bar. With the data provided by GDIGS, we can probe the distribution of the DIG near W43 in more detail than previous observations (see \S \ref{sec:spatial}). We discuss the kinematics of the W43 complex in \S \ref{sec:kinematics} and constrain the \heh ionic abundance ratio of the DIG in \S \ref{sec:abundances}. We analyze the dust temperature distribution in \S \ref{sec:dust} and conclude in \S \ref{sec:conclusions}.

\section{Observations}\label{sec:observations}

GDIGS is a radio recombination line survey at C-band (4--8\,GHz) using the GBT. The average half-power beam width at these frequencies is $\sim 2\arcmin$. We make total-power spectral observations in ``On-the-Fly" (OTF) mode. The telescope is slewed at 54\arcsec\,s$^{-1}$, and data are sampled every 0.38\,s, or 20\arcsec. Rows are spaced every 40\arcsec, which is the Nyquist rate at the highest frequency. We observe a reference position $\sim 3$\degree off the Galactic mid-plane every 16 rows, or $\sim 20$ minutes, such that the time between any on-source scan and a reference scan is less than 10 minutes. To ensure that the reference positions are devoid of RRL emission, we perform a total-power pointed observation on each reference position prior to the mapping observations. 

The pointed observations use the same setup as the OTF maps, with On- and Off-source integration times of 6 minutes per scan. The Off-source scans for these observations track the same azimuth and zenith angle path as the On-source scans, and are offset 6.5 minutes in RA such that they follow the same path on the sky. 

We configure the VEGAS backend to simultaneously tune to 64 different spectral windows at two orthogonal polarizations, each within the 4--8\,GHz receiver bandpass. Each spectral window spans 23.44\,MHz with a spectral resolution of 2.9\,kHz, or a velocity resolution of 0.1--0.2\kms over the receiver bandpass. Of these 64 tunings, 22 are Hn$\alpha$ lines from $n = 95$ to $117$, excluding H113$\alpha$, which is compromised by the nearby H142$\beta$ line. In our configuration, the performance of the C-band receiver is poor below 4.3 GHz and above 7.5 GHz, which typically reduces the number of good Hn$\alpha$ lines to 15 (namely the H97$\alpha$ through H111$\alpha$ transitions). 

The advantage of observing a large number of Hn$\alpha$ transitions simultaneously is that we can average all lines at a given position to make one sensitive spectrum \citep{Balser2006}, a technique that is well-understood \citep{Anderson2011,Liu2013,Alves2015,Luisi2018}. After removing transient RFI, we subtract a polynomial baseline for each transition and shift the spectra so that they are aligned in velocity \citep{Balser2006}. We re-grid the 15 good Hn$\alpha$ lines to a velocity resolution of 0.5\kms and a spatial resolution of 30\arcsec\ using the \emph{gbtgridder}\footnote{https://github.com/GreenBankObservatory/gbtgridder} with a Gaussian kernel. We then average the individual maps using a weighting factor of $t_{\rm intg} T^{-2}_{\rm sys}$ where $t_{\rm intg}$ is the integration time and $T_{\rm sys}$ is the system temperature. We finally perform a 0th-order baseline correction on the averaged maps by subtracting the median value of each single spectrum. To account for possible hydrogen and helium RRLs, we only use the known line-free portion of the spectrum (at hydrogen and helium velocities outside the 10--130\kms range) in calculating the median. 

\begin{figure*}
\centering
\begin{tabular}{cc}
\includegraphics[width=.5\textwidth]{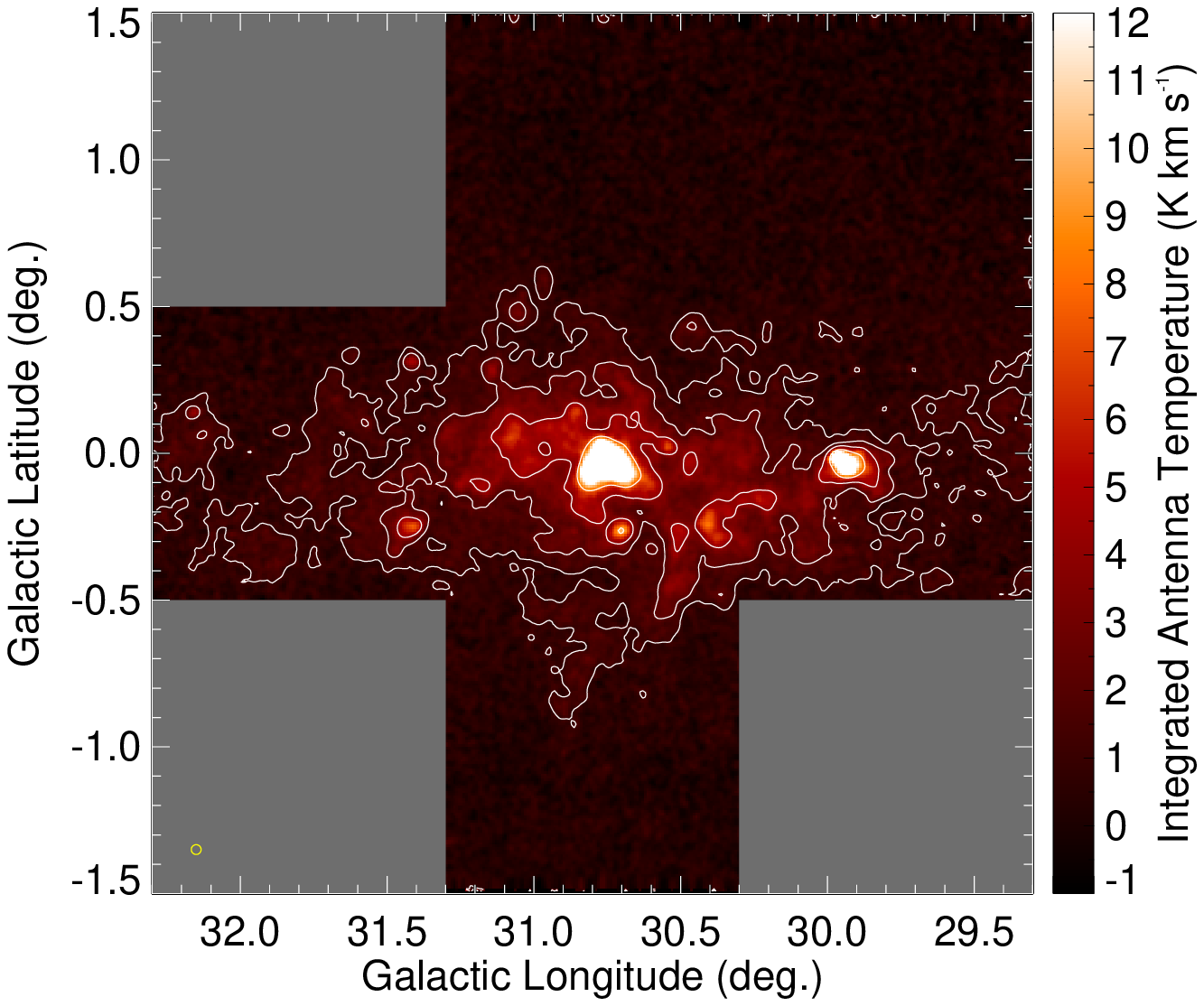}
\includegraphics[width=.5\textwidth]{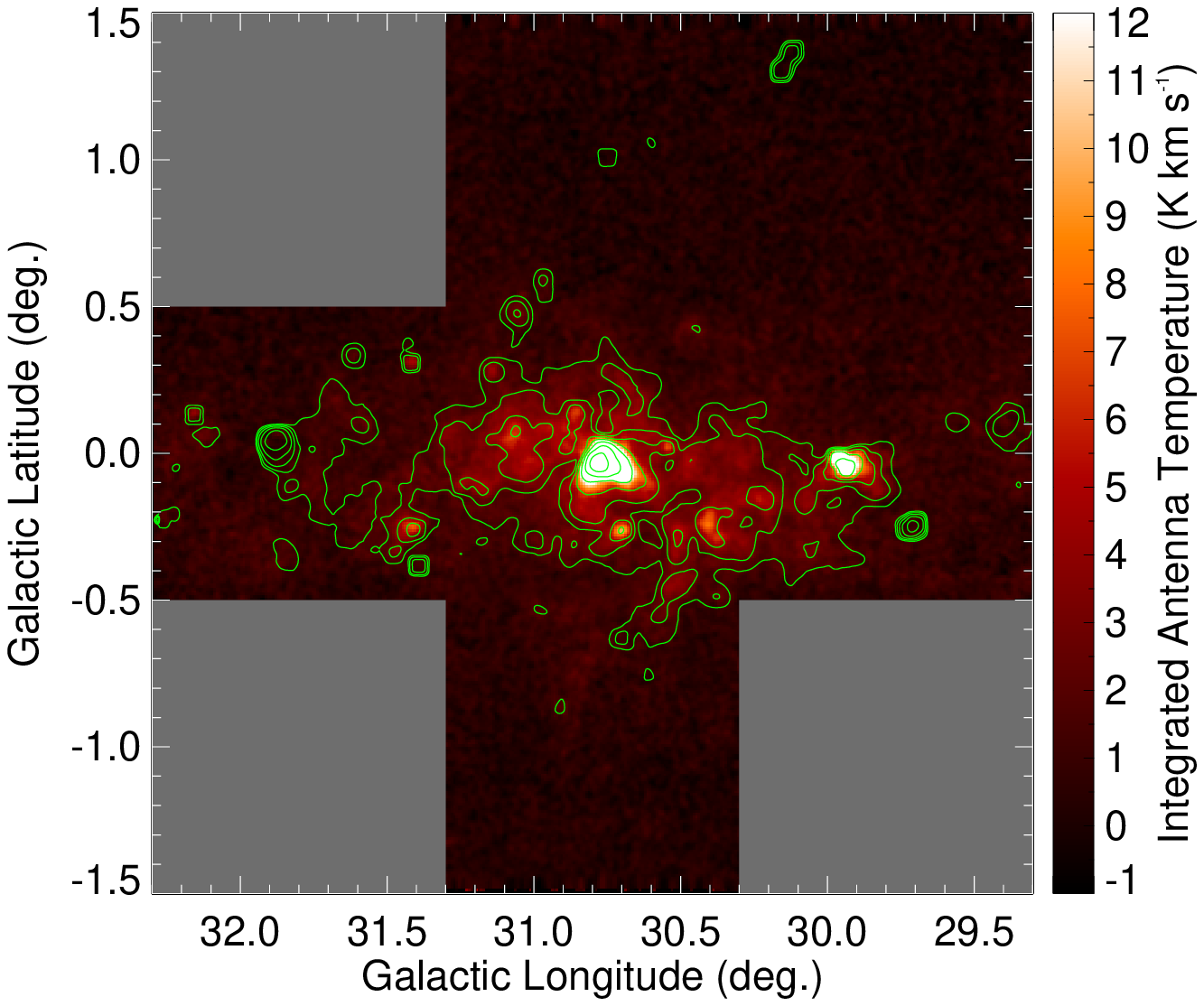}\vspace{-14pt}
\end{tabular}
\caption{Left: Moment 0 map of the W43 complex, integrated from 10 to 130\kms. The average GBT beam size is shown as the yellow circle in the bottom left corner. Contours are at 1, 2, 4, and 8\,K\kms. The rms noise per pixel of the moment 0 map is 0.28\,K\kms. The emission at $\ell \sim 30.8\degree$ is due to W43 itself (W43-Main), while G29 (W43-South) is at $\ell \sim 29.9\degree$. Essentially all visible emission is due to hydrogen. Right: Same, but the green contours are VGPS 1.4\,GHz continuum emission and are logarithmically spaced between 5 and 320\,K. Bright VGPS regions not associated with strong RRL emission are typically supernova remnants \citep[e.g.,][]{Green2014}. \label{fig:mom0_w43}}
\end{figure*}

We initially calibrate the intensity scale of our data using noise diodes fired during data acquisition. While the derived flux densities of the averaged maps agree with the values for the continuum flux calibrator 3C286 \citep[see][]{Ott1994} to within 4\% regardless of elevation and weather conditions, the flux densities of the individual spectral windows may show deviations up to $\sim 10$\%. We therefore apply an intensity correction based on periodic mapping observations of the primary flux calibrator 3C286 to all individual maps prior to averaging. This reduces the deviation in intensity of the individual maps to $<3$\%. We also periodically verify the calibration scale by observing the RRL emission from W43 using pointed observations with the same setup. In our map, the root mean square (rms) noise per pixel within a single 0.5\kms velocity channel is 16.0\,mK, which corresponds to 4.8\,mK\,channel$^{-1}$ or 2.4\,mJy\,beam$^{-1}$\,channel$^{-1}$. When smoothed to a spectral resolution of 5\kms, the rms noise is 0.78\,mJy\,beam$^{-1}$.

The GDIGS data discussed here encompass a total of six square degrees surrounding the \hii\ region W43. Our maps include three square degrees within the Galactic plane from $32.3\degree > \ell > 29.3\degree$, $|b| < 0.5\degree$, as well as a two square degree exploration above the plane ($31.3\degree > \ell > 29.3\degree$, $0.5\degree < b < 1.5\degree$) and one square degree below the plane ($31.3\degree > \ell > 30.3\degree$, $-1.5\degree < b < -0.5\degree$).

\section{Spatial Distribution of the DIG near W43}\label{sec:spatial}
The W43 complex contains two large regions of active star formation: W43-Main at $\ell \sim 30.8\degree$ and W43-South at $\ell \sim 29.9\degree$. W43-Main (hereon simply referred to as ``W43") is one of the brightest radio sources in the sky, with a GBT peak intensity of $\sim 2$\,K ($\sim 1$\,Jy) at X-band \citep{Balser2011} and a total IR luminosity of $\sim 6.5 \times 10^6$\lsun \citep[see][]{Blum1999}. In W43's core lies a giant \hii\ region ionized by a cluster of Wolf-Rayet and OB stars. W43-South, here referred to as ``G29", harbors the compact \hii\ region G29.96--0.02, which has an IR luminosity of $\sim 1.4 \times 10^6$\lsun \citep{Cesaroni1994}. There are numerous other \hii\ regions in the vicinity of the W43 complex. In the \emph{WISE} Catalog of Galactic \hii\ regions \citep[V2.2;][]{Anderson2014}, 298 \hii\ regions and \hii\ region candidates lie within the six square degrees mapped. Of these 298 regions, 115 are within 30\arcmin\ of W43 and 76 are within 30\arcmin\ of G29.

Due to the large luminosities and the large number of discrete \hii\ regions near W43 and G29, DIG emission should be strong toward these directions. As part of a study of the compact \hii\ region NGC\,7528, \citet{Luisi2016} discovered diffuse hydrogen RRL emission near W43 with GBT X-band antenna temperatures $> 10$\,mK up to several hundred pc away from the ionizing star cluster. In a follow-up study of the large-scale distribution of the DIG, \citet{Luisi2017} confirmed that there is strong large-scale diffuse RRL emission across several square degrees near $\sim 30\degree$. Due to the coarse resolution, however, the spatial correlation between \hii\ regions and the DIG could not be probed in detail. Below we show our recent GDIGS observations of the W43 complex and examine the relationship between RRL emission from discrete \hii\ regions and diffuse RRL emission.

\subsection{GDIGS mapping data}\label{sec:mappingdata}
There are numerous bright RRL sources in the GDIGS datacube. As expected, the sources with the largest hydrogen RRL intensities are W43 and G29. We show a moment 0 map of the surveyed area in Figure~\ref{fig:mom0_w43}. In order to maximize our signal-to-noise (S/N) ratio, the moment 0 map is integrated over the 10--130\kms local standard of rest (LSR) velocity range, within which we expect nearly all hydrogen RRL emission (see \S \ref{sec:kinematics}). We find that 41 of the 298 \hii\ regions in the map have an integrated hydrogen emission intensity $> 10$\,K\kms and four have integrated emission $> 100$\,K\kms.

We find an angularly large low-intensity structure below the Galactic plane that is clearly visible in the moment 0 map (Figure~\ref{fig:mom0_w43}, left panel). This structure extends downward from W43 to $b \sim -1\degree$ where the intensity drops below the detection limit. We do not detect a counterpart to this structure above the plane. The structure may be associated with the ``worm-ionized medium" \citep[e.g.,][]{Heiles1996a}, low-density ionized gas within filamentary structures extending outwards from the Galactic plane. The worms may act as pathways for Lyman continuum photons to escape from the mid-plane. The structure in Figure~\ref{fig:mom0_w43} coincides spatially and in velocity with the worm ``GW30.5--2.5" found by \citet{Heiles1996a}. Based on their S-band continuum data, GW30.5--2.5 extends from W43 to $b \sim -4\degree$, and from $\ell \sim 30\degree$ to 32\degree. It is likely that here we detect only the denser part of GW30.5--2.5. The fact that there are no discrete \hii\ regions near $\ell \approx 30.9\degree$, $b \approx -1.0\degree$ (see \S \ref{sec:wimonly}) suggests that the majority of emission within this structure is truly due to the DIG.

From our GDIGS maps, we can constrain the emission measure (EM), 
\begin{equation}
    \frac{\textnormal{EM}}{\textnormal{pc\,cm}^{-6}} \equiv \int _{\rm los} \left( \frac{n_{\rm e}}{\textnormal{cm}^{-3}} \right) ^2 d \left( \frac{s}{\textnormal{pc}} \right),
    \label{eq:em1}
\end{equation}
where $n_{\rm e}$ is the electron number density and $s$ is the path length, integrated along the line of sight. For Hn$\alpha$ lines and assuming an observing frequency of 5912.5\mhz,
\begin{equation}
    \frac{\textnormal{EM}}{\textnormal{pc\,cm}^{-6}} \approx 1.01\times10^{-2}\,\left( \frac{T_{\rm L}}{\textnormal{K}} \right) \left( \frac{\Delta V}{\textnormal{\kms}} \right) \left( \frac{T_{\rm e}}{\textnormal{K}} \right)^{3/2},
    \label{eq:em2}
\end{equation}
where $T_{\rm L}$ is the hydrogen RRL intensity, $\Delta V$ is the hydrogen RRL full width at half maximum (FWHM), and $T_{\rm e}$ is the electron temperature \citep[see][Anderson et al., in prep.]{Condon2016}. Using Equation~\ref{eq:em2} and assuming $T_{\rm e} = 8000$\,K, we find a maximum EM of $\sim 2 \times 10^5$\,pc\,cm$^{-6}$ toward W43. The rms (per-pixel) sensitivity of our data is $\textnormal{EM} \approx 2000$\,pc\,cm$^{-6}$.

\begin{figure}
\centering
\includegraphics[width=0.48\textwidth]{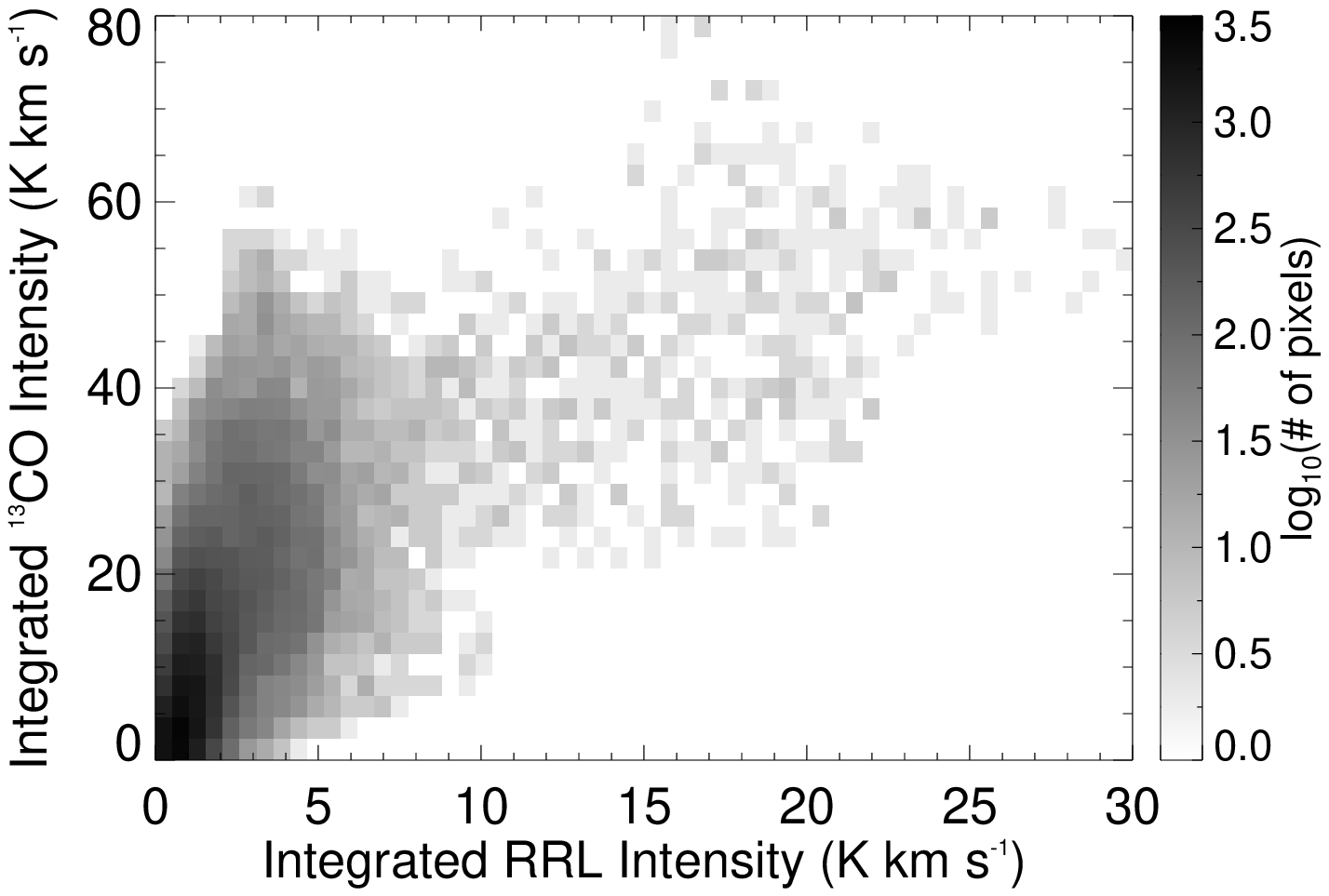}\\
\includegraphics[width=0.48\textwidth]{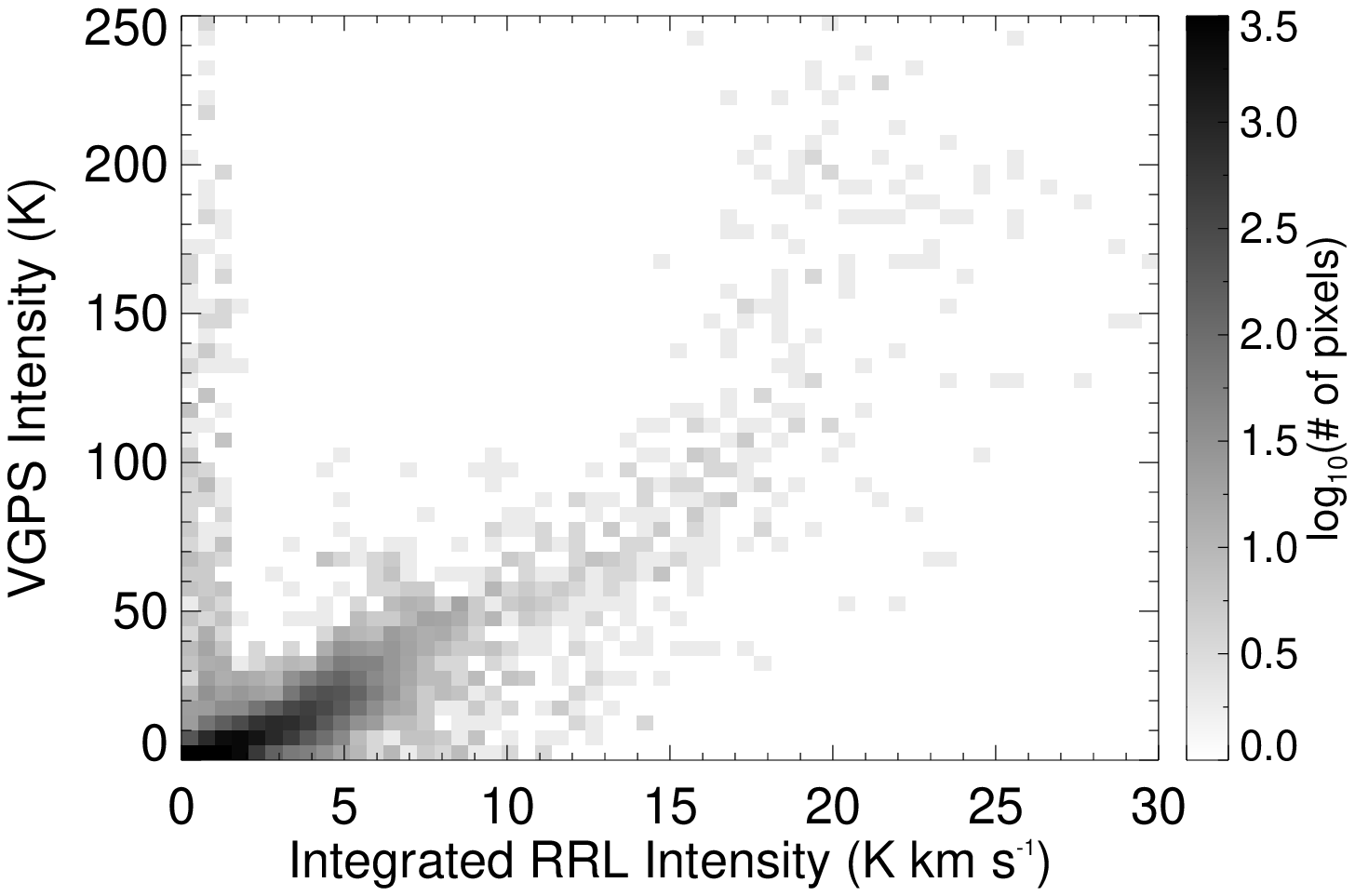}
\caption{Top: Density plot of the pixel-by-pixel correlation between the integrated intensity GDIGS RRL map and GRS $^{13}$CO data. The grayscale represents the number of pixels within each grid point. The GRS data are gridded to the lower resolution of the GDIGS data. There is a strong correlation between RRL intensity and $^{13}$CO emission ($r = 0.73$). Bottom: Correlation between the GDIGS RRL map and 1.4\,GHz VGPS data ($r = 0.67$). The data points with large VGPS intensity but weak RRL emission are mostly due to supernova remnants and active galactic nuclei. Since the GBT beam is larger than the pixel size, the intensities of pixels within one beam area are not truly independent of each other. \label{fig:corr_co}}
\end{figure}

Comparison with other gas tracers shows that there is a strong correlation between RRL intensity and $^{13}$CO emission from the Galactic Ring Survey \citep[GRS;][]{Jackson2006} tracing the denser molecular gas. We show the pixel-by-pixel correlation after regridding the GRS data to the resolution of the GDIGS RRL map in the top panel of Figure~\ref{fig:corr_co}. We calculate Pearson's $r$ as a measure of the correlation and find $r = 0.73$. The large value of $r$ is expected since the W43 complex is one of the most active regions of ongoing star formation.

There is also a correlation ($r = 0.67$) between RRL intensity and 1.4\,GHz radio continuum emission from the Very Large Array (VLA) Galactic Plane Survey \citep[VGPS;][see bottom panel of Figure~\ref{fig:corr_co}]{Stil2006}, a 1\arcmin-resolution, 1.4\,GHz VLA survey with additional short-spacing data from the 100\,m Effelsberg telescope \citep{Reich1986,Reich1990}.

Numerous sources are bright in 1.4\,GHz continuum but not in RRLs; these regions are predominantly supernova remnants \citep{Green2014} and active galactic nuclei (see Figure~\ref{fig:mom0_w43}, right panel). The continuum intensity of these sources is dominated by synchrotron emission. The change in slope at integrated RRL intensities above $\sim 18$\,K\,km\,s$^{-1}$ in the bottom panel of Figure \ref{fig:corr_co} may indicate increased non-thermal emission in the regions of strongest RRL emission.

The correlation between RRL intensity and H$\alpha$ emission from the WHAM Sky Survey \citep[][not shown in the figure]{Haffner2003} is poor ($r = 0.24$), presumably because the spatial resolution of WHAM is only $\sim 1\degree$ and because the W43 complex is known to be optically obscured \citep[e.g.,][]{Green2019}.

\subsection{DIG-only data}\label{sec:wimonly}
In order to study the diffuse gas emission unaffected by emission from discrete \hii\ regions, it is necessary to disentangle the RRL emission based on its origin. We know the locations and angular sizes of all \hii\ regions and \hii\ region candidates from the \emph{WISE} catalog and can blank emission from within these regions. The \emph{WISE} region sizes are determined by their IR morphology: 22\,$\mu$m emission from hot dust grains within the ionized \hii\ region volume surrounded by 12\,$\mu$m emission from PAHs tracing the \hii\ region PDR. Since PAHs cannot survive in a strong radiation field, the \emph{WISE} region radii determined by the 12\,$\mu$m emission usually overestimate the size of the fully ionized \hii\ regions. Without knowledge of the \hii\ region velocities, however, we are forced to blank the GDIGS data across all spectral channels. This possibly results in the loss of diffuse emission, since diffuse gas may lie along the same line of sight as the \hii\ region. Fortunately, the \emph{WISE} catalog includes velocity information for 99 confirmed \hii\ regions within the mapped range \citep[partially compiled by the HRDS; see][]{Bania2012}. For these regions, we only exclude data within $\pm 1.5\,{\rm FWHM}$ of the \hii\ region velocity, whereas we blank all velocity channels for the remaining 198 \hii\ region candidates.

\begin{figure}
\centering
\includegraphics[width=0.5\textwidth]{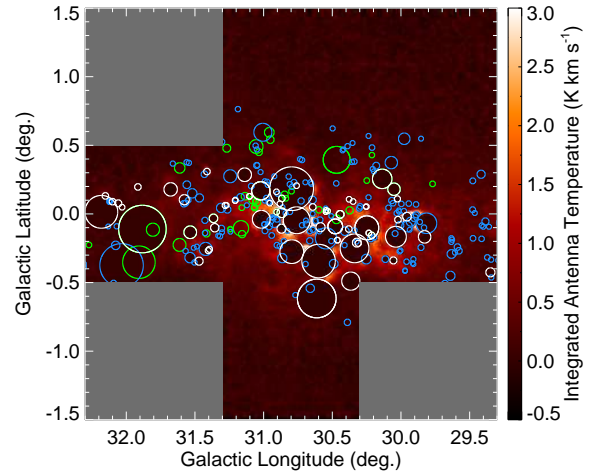}
\caption{Diffuse RRL emission, integrated from 90\kms to 110\kms. Emission from known \hii\ regions within these velocities (white circles) was cut from the map. We also removed emission from \hii\ region candidates (blue circles). The green circles indicate \hii\ regions outside the 90\kms to 110\kms velocity range. \label{fig:100kms_w43_masked}}
\end{figure}

\begin{figure}
\centering
\includegraphics[width=0.5\textwidth]{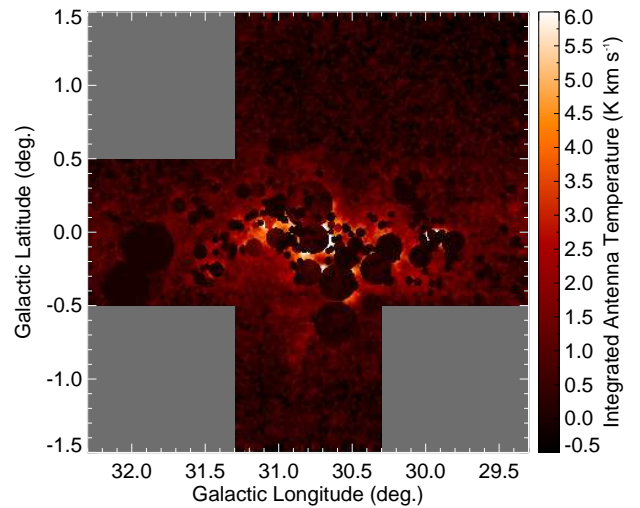}\vspace{-10pt}
\caption{DIG-only moment 0 map. Velocity channels within $\pm 1.5 \times {\rm FWHM}$ of the \hii\ region velocity were blanked for each known \hii\ region. For \hii\ region candidates without known velocities, we removed emission from all velocity channels. There is extended DIG emission between W43 and G29. \label{fig:mom0_w43_masked}}
\end{figure}

As an example, we show the W43 complex integrated over the $90-110$\kms velocity range with \hii\ regions from the \emph{WISE} catalog displayed and blanked (Figure~\ref{fig:100kms_w43_masked}). We also show the \hii\ region-subtracted moment 0 map, or ``DIG-only" map, in Figure~\ref{fig:mom0_w43_masked}. We detect diffuse RRL emission across essentially the entire survey range, although it is faint at $|b| > 1\degree$. The total integrated hydrogen intensity between 10 and 130\kms is $2{,}772$\,K\kms for the DIG-only data, compared to $5{,}096$\,K\kms for the unmasked datacube. Even if our measurements of some \hii\ region sizes and velocities are inaccurate, the vast majority of the emission shown in Figure~\ref{fig:mom0_w43_masked} cannot be explained by \hii\ region emission alone. We find the strongest DIG emission near W43 and in the zone between W43 and G29, suggesting that at these locations the ISM is strongly affected by radiation leaking from \hii\ regions. We note that the brightest pixels of the DIG-only map are located directly outside W43 and may trace RRL emission from the \hii\ region itself, possibly because the radius of W43 is underestimated in the \emph{WISE} catalog.

We derive the average scale height, $h$, of the DIG-only RRL data at all longitudes where our maps extend to $|b| = 1.5\degree$, and find $h = 30 \pm 4\arcmin$. Assuming a distance of 6\,kpc (see \S \ref{sec:kinematics}), this corresponds to $h = 52 \pm 7$\,pc. This number is much lower than that derived from the combination of pulsar and H$\alpha$ data \citep[$250 < h < 500$\,pc;][]{Berkhuijsen2006}, suggesting that we are sensitive to only the denser DIG components near discrete \hii\ regions.

\subsection{Modeling the DIG Emission}\label{sec:model}
If the ionization of the DIG is maintained by O-type stars within \hii\ regions, it should be possible to estimate the spatial distribution of the DIG from the \hii\ region distribution. The \emph{WISE} catalog provides an ideal tool for this analysis as it is essentially complete in the inner Galaxy and contains the locations and angular sizes of all \hii\ region candidates. 

\begin{deluxetable*}{llccccc}
\tablewidth{0pt}
\tablecaption{\hii\ Region Model Parameters \label{tab:model}}
\tablehead{\colhead{Velocity range} & \colhead{Functional form} & \colhead{$k$} & \colhead{$m$} & \colhead{Pearson's $r$} & \colhead{$\sum \epsilon ^2$}\\
\colhead{(\kms)} & \colhead{} & \colhead{} & \colhead{} & \colhead{} & \colhead{(K$^2$\,km$^2$\,s$^{-2}$)} }
\startdata
10$-$130 & Exponential & $0.133 \pm 0.033$ & $0.33 \pm 0.05$ & 0.859 & 20{,}627\\
10$-$130 & Power law   & $0.262 \pm 0.071$ & $1.83 \pm 0.11$ & 0.880 & 17{,}452\\
\hline
10$-$65  & Exponential & $0.165 \pm 0.041$ & $0.32 \pm 0.05$ & 0.677 & \phn 3{,}795\\
10$-$65  & Power law   & $0.307 \pm 0.083$ & $1.81 \pm 0.11$ & 0.683 & \phn 3{,}676\\
\hline
65$-$130 & Exponential & $0.148 \pm 0.037$ & $0.36 \pm 0.05$ & 0.879 & 11{,}456\\
65$-$130 & Power law   & $0.276 \pm 0.075$ & $1.89 \pm 0.12$ & 0.905 & \phn 8{,}917
\enddata
\end{deluxetable*}

Photoionization models \citep[e.g., CLOUDY;][]{Ferland2017} can, in principle, predict the ionizing radiation field around a single \hii\ region given enough information on that region. Galactic \hii\ regions, however, vary substantially in morphology and in their physical parameters (e.g., density), and for the majority of \hii\ regions in our survey these parameters are unknown. It is therefore unfeasible to model each \hii\ region based on its individual properties. Instead, we aim to estimate the large-scale distribution and intensity of the DIG from a purely empirical model with as few input parameters as necessary. We restrict our model inputs to only the location and size of each \hii\ region, and its average GDIGS RRL intensity.

In a study of eight Galactic \hii\ regions, \citet{Luisi2019} discovered that the slope of the hydrogen RRL intensity with distance from the \hii\ region is roughly inversely proportional to the \hii\ region size traced by its PDR. This is in agreement with previous studies \citep[e.g.,][]{Zurita2000}, who found that large \hii\ region complexes are believed to affect the ISM out to larger distances compared to compact \hii\ regions. For \hii\ regions in a constant density environment, this discovery is supported by simple geometry. The RRL emission intensity is proportional to the number density squared and therefore this relationship does not hold for variations in density, since high-density ionized gas will generate disproportionately large RRL intensities compared to low-density gas. For our model, however, we assume that that the angular size of the \hii\ region constrains the slope of RRL emission with distance from that region.

The functional form of the hydrogen RRL emission intensity with distance from an \hii\ region is not well-understood. Previous work showed that the hydrogen RRL emission decreases roughly as a power-law beyond the \hii\ region PDR \citep{Luisi2016} for the two studied \hii\ regions: NGC\,7538 and W43. The difference between the power-law model and an exponential function, however, was marginal. There is no statistically significant difference between these two models in the RRL emission surrounding the eight Galactic \hii\ regions observed in \citet{Luisi2019}. 

In our model we assume that all DIG emission is due to photons leaking from \hii\ regions and that the diffuse hydrogen RRL emission decreases with distance from an \hii\ region using two different functional forms: an exponential (see \S \ref{sec:exp}) and a power law (see \S \ref{sec:pl}). The model treats each region as perfectly circular and isotropic, with a radius equal to the radius given in the \emph{WISE} catalog. We further assume that the RRL emission intensity at the boundary of each \hii\ region is proportional to the average integrated RRL intensity within the region. We derive the average integrated intensities from our GDIGS moment 0 map, using the \hii region sizes defined by the \emph{WISE} catalog.

We compare each model with the GDIGS DIG-only map and find the best-fit model by minimizing the sum of the squares of the residuals, $\sum \epsilon ^2$, using the GDIGS moment 0 map as the reference. We also compute Pearson's correlation coefficient, $r$, as a proxy for the quality of the model.

\subsubsection{Exponential model}\label{sec:exp}
\begin{figure*}
\centering
\begin{tabular}{cc}
\includegraphics[width=.5\textwidth]{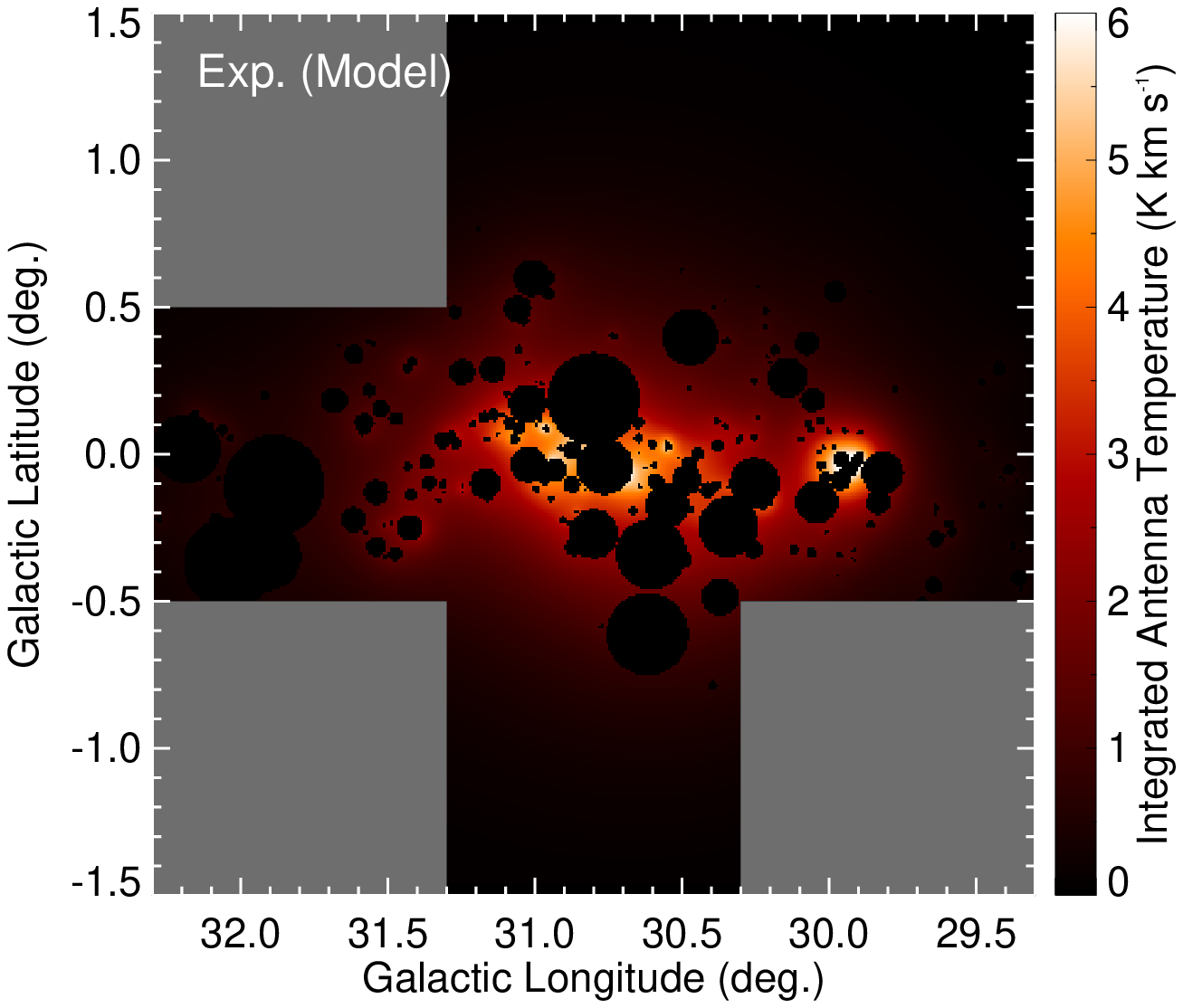} & 
\includegraphics[width=.5\textwidth]{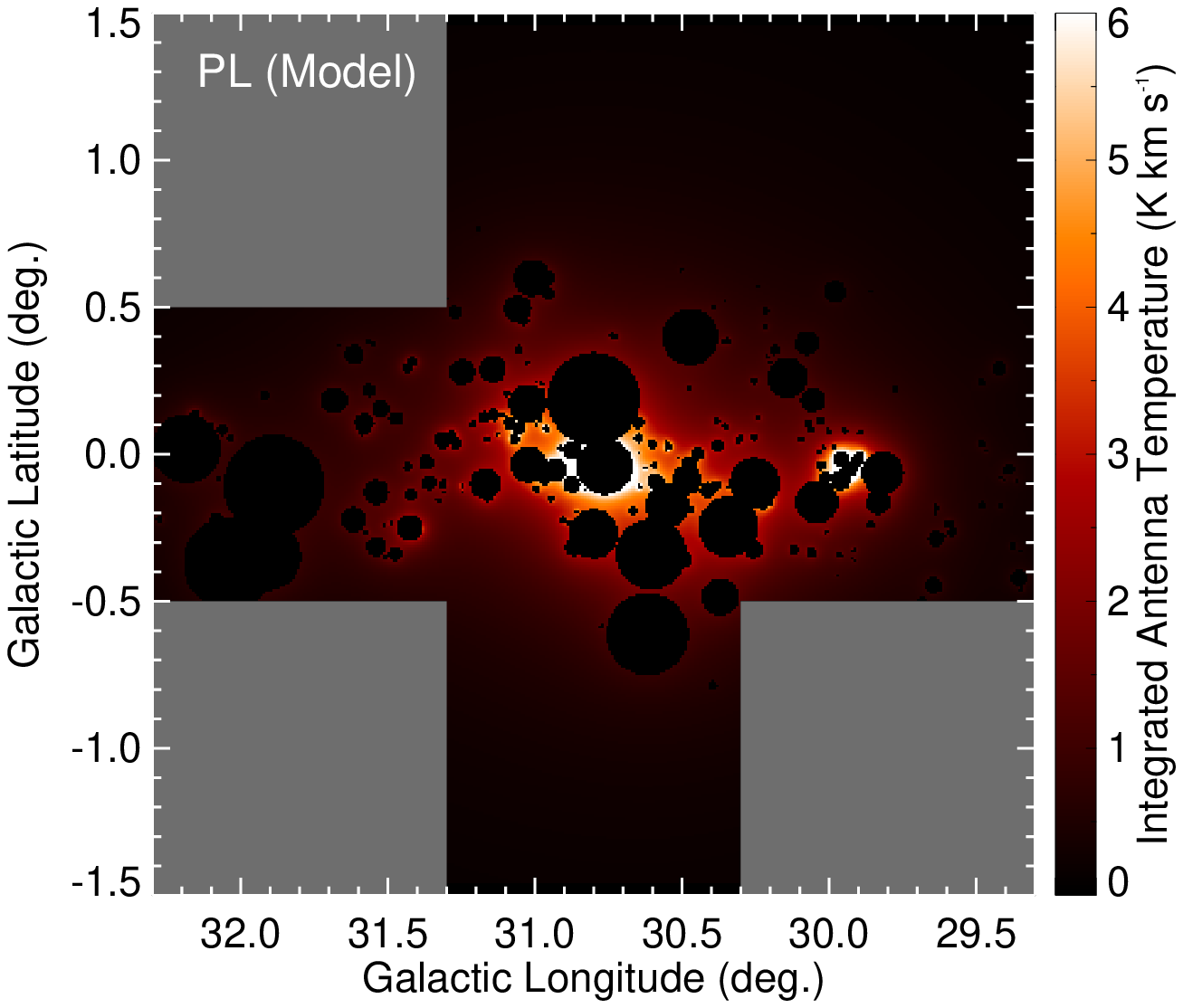}\vspace{-12pt}\\
\includegraphics[width=.5\textwidth]{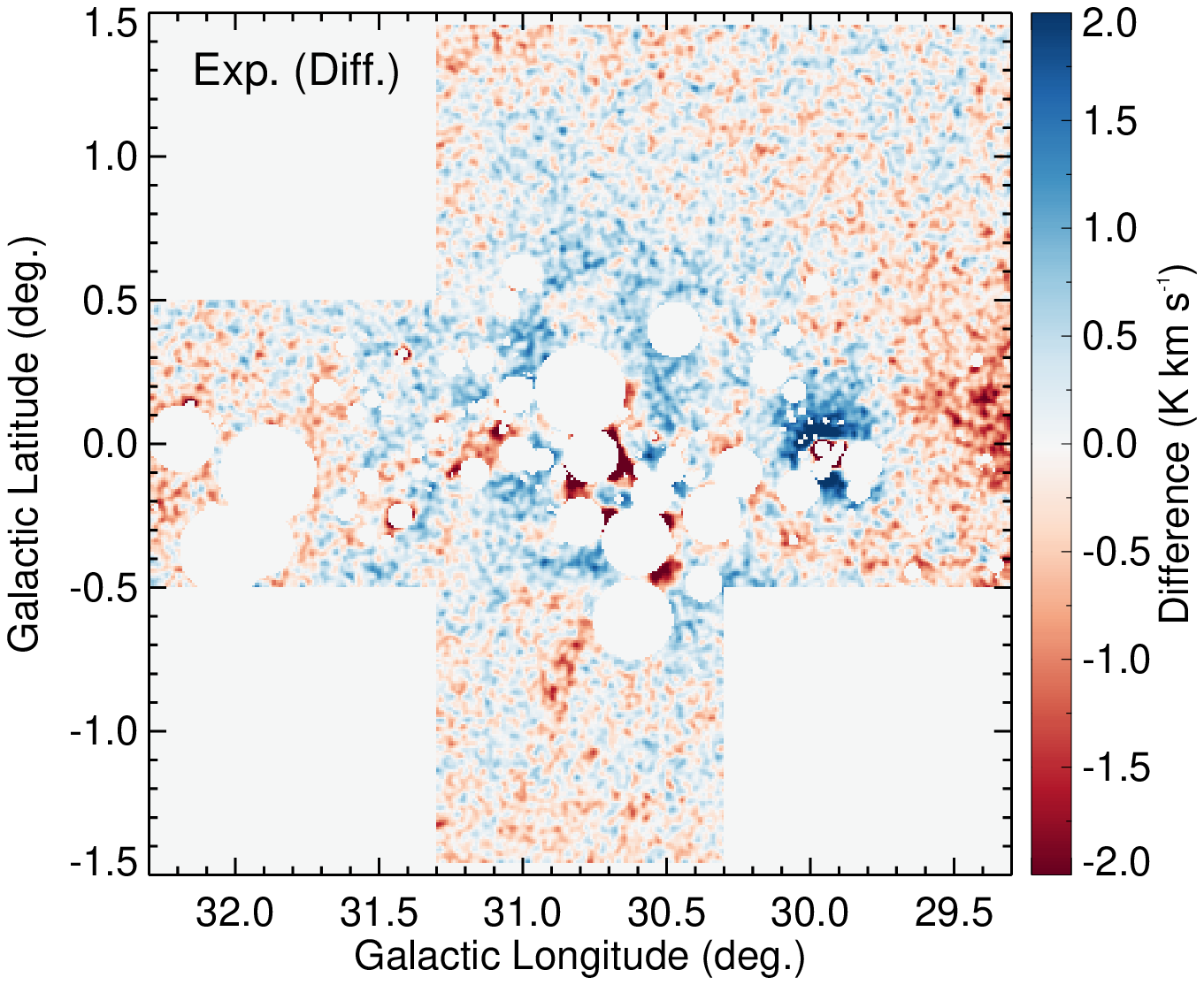} &
\includegraphics[width=.5\textwidth]{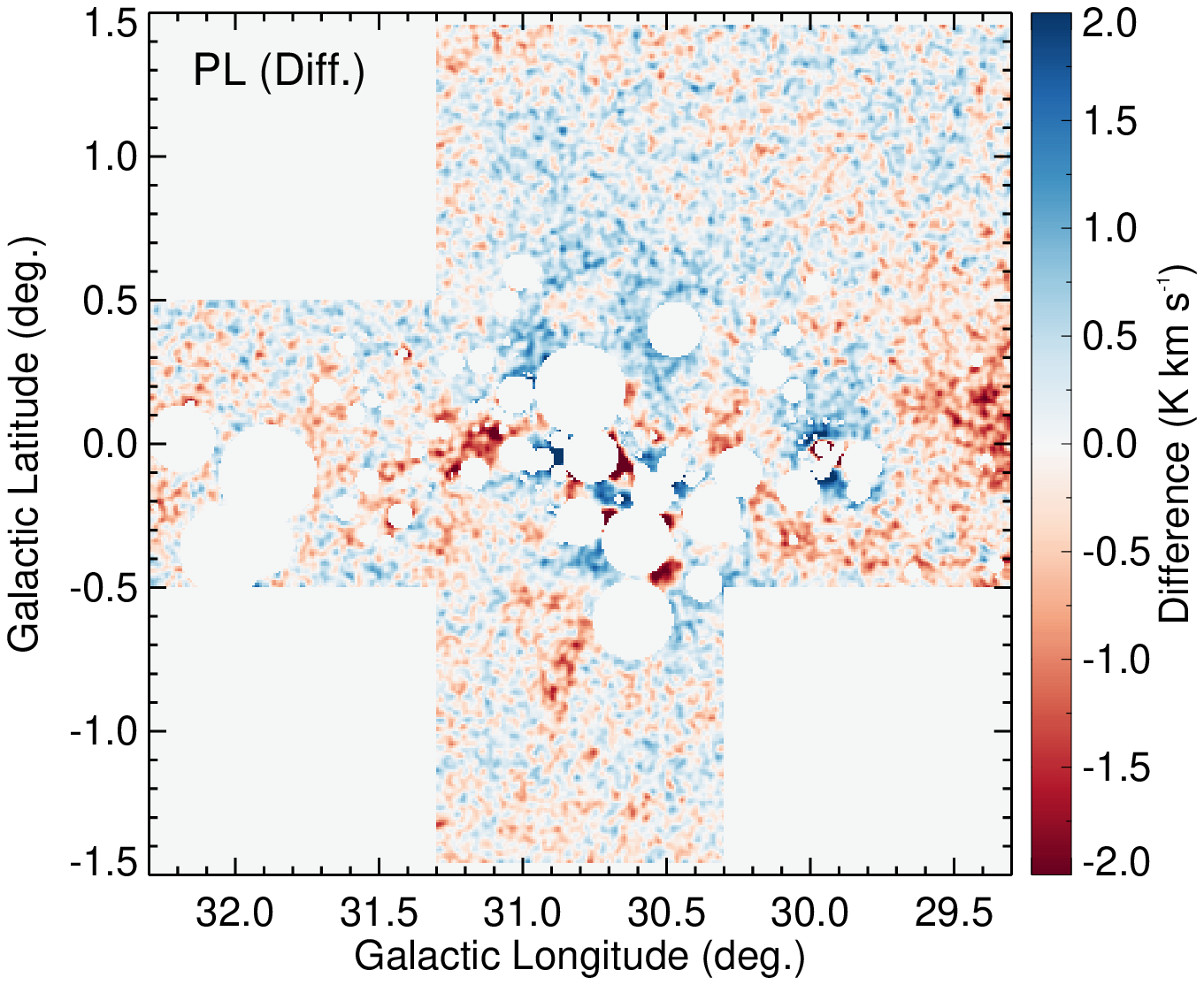}\vspace{-12pt}
\end{tabular}
\caption{Top left: model of diffuse hydrogen RRL emission, assuming an exponential decrease in RRL intensity with distance from each \hii\ region. Bottom left: difference between the GDIGS moment 0 data and the model. Larger values indicate stronger emission in the model. Top right: same as top left, but assuming a power-law decrease in RRL emission with distance from each \hii\ region. The power-law model is a better fit to the data (see \S \ref{sec:model}). Bottom right: difference between the GDIGS moment 0 data and the power-law model. \label{fig:model}} 
\end{figure*}

For the exponential model, we assume
\begin{equation}
T_{\rm A, \, model}(x,y) = k \sum_{i=1}^{N} T_{{\rm A}, \, i} \, \exp \left( -m \frac{r_i}{r_{{\rm PDR}, \, i}} \right),
\end{equation}
where $T_{\rm A, \, model}$ is the antenna temperature of the model for each individual pixel in the map, $k$ and $m$ are the fit parameters, $N=298$ is the total number of \hii\ regions in the map, $T_{{\rm A,} \, i}$ is the GDIGS moment 0 map intensity averaged over \hii\ region $i$, $r_i$ is the distance from the \hii\ region, and $r_{{\rm PDR,} \, i}$ is the radius of the region as given in the \emph{WISE} catalog. The parameter $k$ is always calculated such that the total observed flux is reproduced, i.e.~the total integrated intensity of the model map must be equal to the total integrated intensity of the moment 0 map; $k$ is therefore not truly a free parameter. We give our best-fit values for $k$, $m$, Pearson's $r$, and the sum of the squares of the residuals, $\sum \epsilon ^2$, in Table~\ref{tab:model}. We show the model in the left panel of Figure~\ref{fig:model}.

We assume that the major uncertainty contributions for the model are the \hii\ region sizes defined by the \emph{WISE} catalog and the averaged integrated intensities from our moment 0 map; we assume a 20\% uncertainty for both. We perform a Monte Carlo uncertainty estimation of the model output parameters $k$ and $m$ by resampling the input parameters $T_{{\rm A,} \, i}$ and $r_{{\rm PDR,} \, i}$ from a Gaussian probability distribution centered on their nominal values and a standard deviation of 20\%. We recompute $k$ and $m$ ten times (since the model is computationally intensive, we are restricted in the total number of samples). The uncertainty values given in Table~\ref{tab:model} are the standard deviations of the recomputed values of $k$ and $m$.

An additional source of uncertainty is the rms noise in the GDIGS moment 0 map, as well as the fact that real \hii\ regions are neither homogeneous nor spherically symmetric. By adding random noise to the GDIGS map and recomputing $k$ and $m$, we find that the rms noise of the map has a negligible impact on the overall uncertainty ($<1$\%). Since we cannot model each \hii\ region separately, the uncertainty associated with the \hii\ region morphologies is unknown. Based on visual inspection of the difference maps between the model and the data, however, the uncertainty contribution must be significant. We therefore consider the uncertainties shown in Table~\ref{tab:model} to be lower limits of the overall model uncertainty.\\

\subsubsection{Power law model}\label{sec:pl}
Here we use a power-law model of the form
\begin{equation}
T_{\rm A, \, model}(x,y) = k \sum_{i=1}^{N} T_{{\rm A}, \, i} \left( \frac{r_i}{r_{{\rm PDR}, \, i}} \right) ^{-m}
\end{equation}
to model the hydrogen RRL emission around W43. Again, we give our best-fit values in Table~\ref{tab:model} and show the model map in the right panel of Figure~\ref{fig:model}.

The power law model is a significantly better fit to the data than the exponential model, both in terms of the squared residuals and Pearson's $r$. The diffuse emission at $\ell>31.5$\degree is marginally brighter in the moment 0 RRL data than either model map, as well as emission at $\ell < 29.8$\degree. This excess emission may indicate that there is a truly diffuse component associated with the Galactic plane that we are not fitting in the model. With the exception of the DIG feature below the plane near $\ell \approx 30.9$\degree, both models estimate the low-intensity DIG at $|b| > 0.5$\degree quite well.  While the exponential model tends to underestimate the RRL intensity near W43 and G29, however, the power law model better represents the bright DIG emission outside these luminous \hii\ region complexes. This difference is particularly noticeable toward the Galactic north of G29. Overall, we find that the model provides a reasonable estimate of the diffuse RRL emission by only considering the aggregate effect of \hii\ regions on the ISM.\vspace{10pt}

\begin{figure}
\centering
\includegraphics[width=0.49\textwidth]{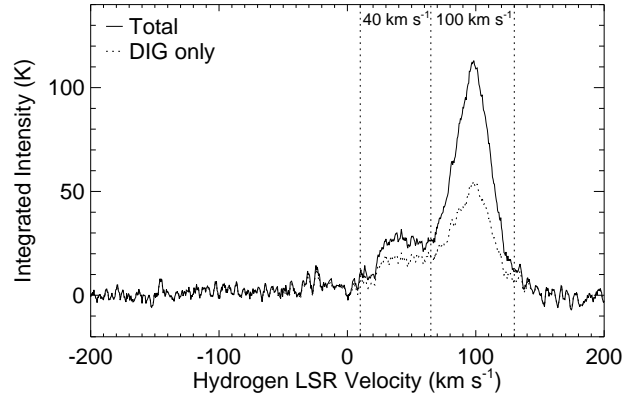}
\caption{RRL spectrum of the mapped area. The emission is mostly found near two velocities: $\sim 40\kms$ and $\sim 100\kms$. Emission from the DIG (dotted line) follows that of the overall ionized gas. The weak emission near a hydrogen LSR velocity of $-25\kms$ is due to helium associated with the $\sim 100\kms$ hydrogen RRL emission. \label{fig:spectrum_w43}}
\end{figure}

\section{Kinematics of the DIG near W43}\label{sec:kinematics}
Due to its location near the end of the Galactic bar, the gas dynamics near W43 are complex. The DIG emission at $\sim 30\degree$ is mainly found at two distinct velocities, $\sim 40\kms$ and $\sim 100\kms$. Determining the distance to the DIG is hampered by not only the kinematic distance ambiguity \citep[KDA; see][]{Anderson2012a}, but also possibly strong streaming motions at the bar/spiral arm interface \citep[e.g.,][]{Beuther2012}. 

\begin{figure*}
\centering
\includegraphics[width=0.65\textwidth]{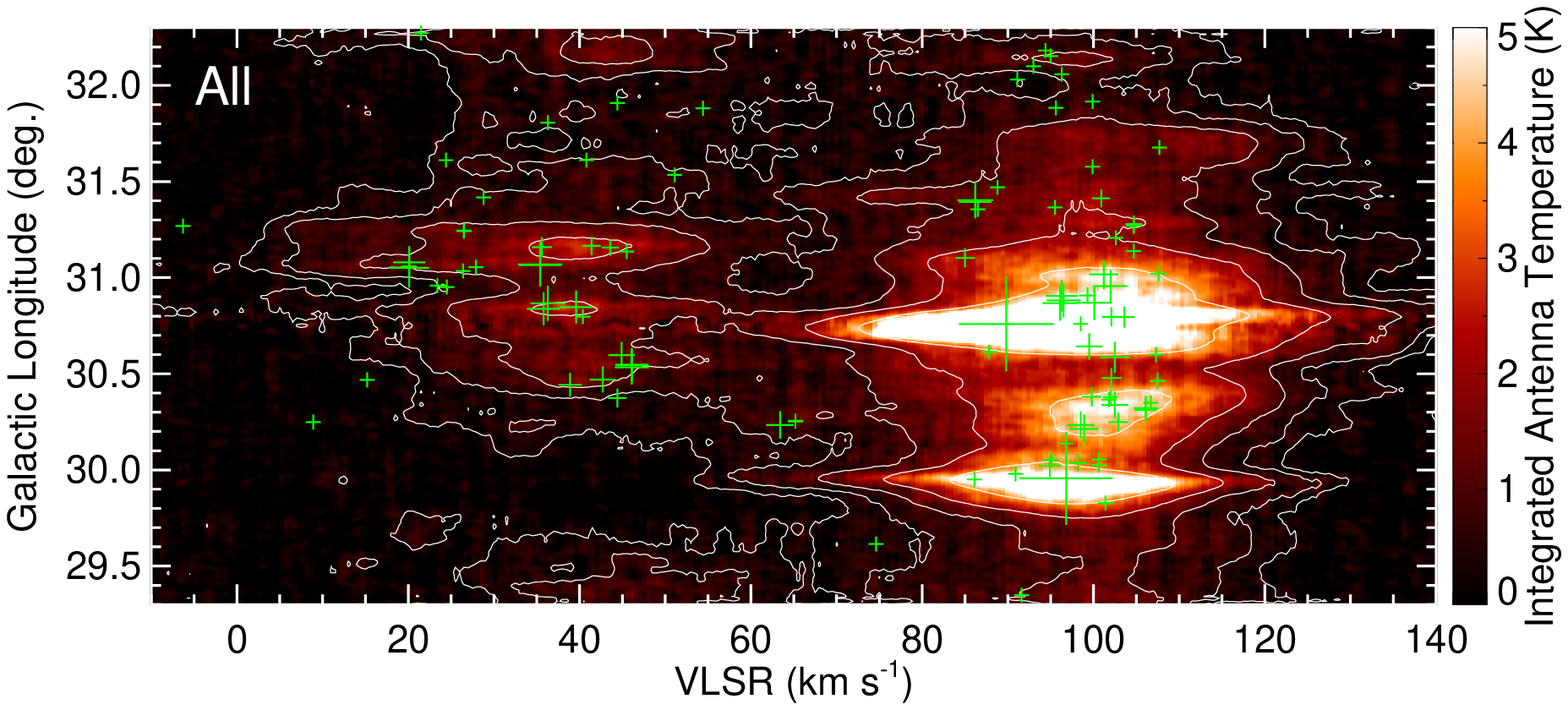}\\
\includegraphics[width=0.65\textwidth]{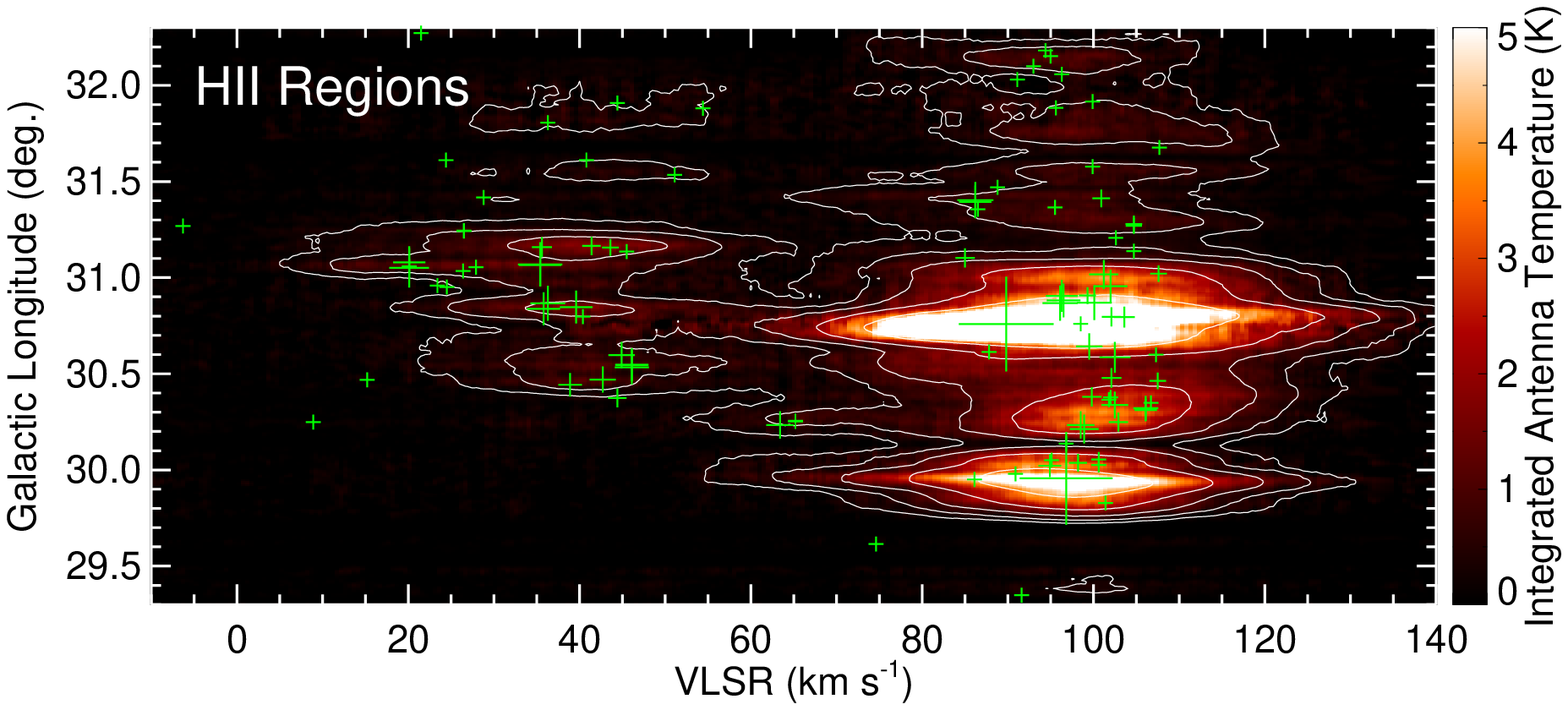}\\
\includegraphics[width=0.65\textwidth]{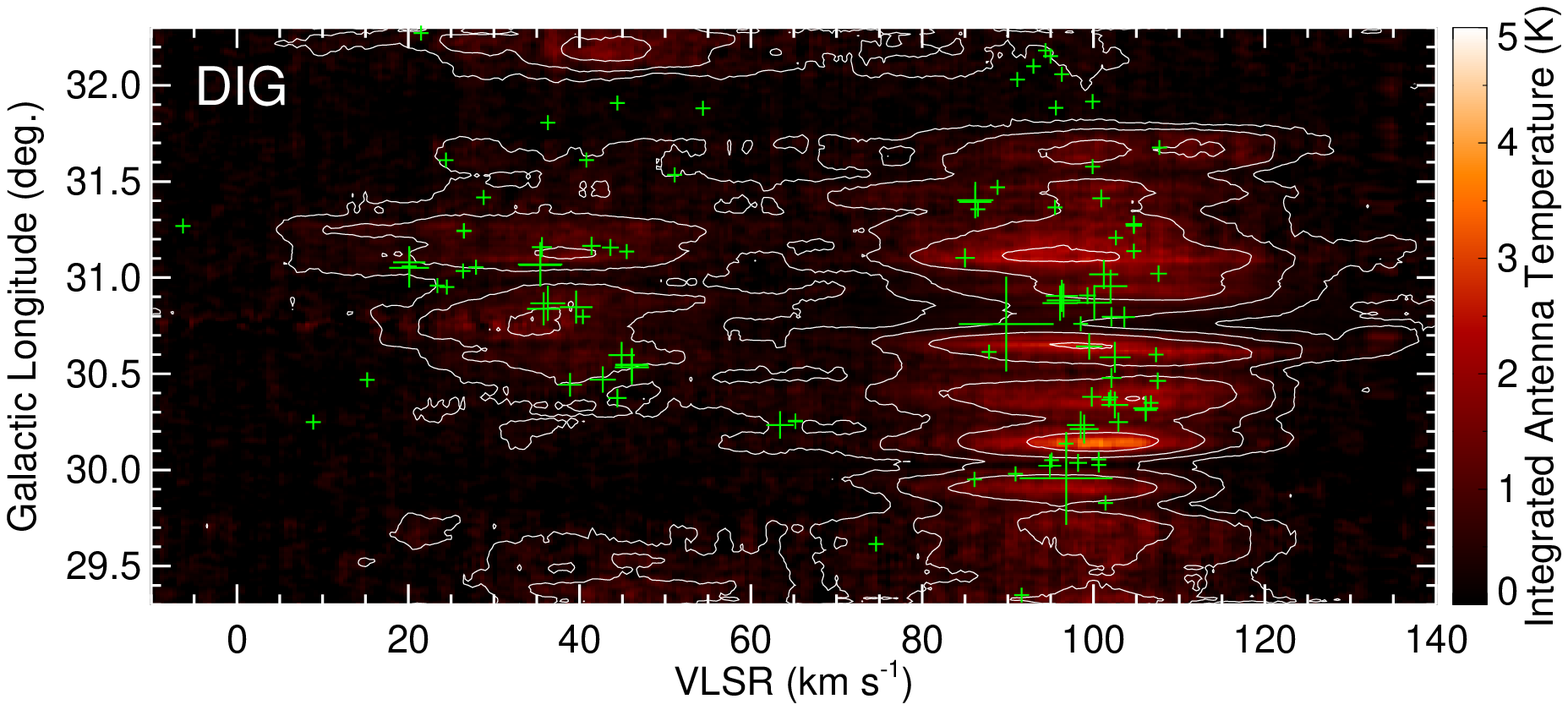}
\caption{Top: longitude--velocity diagram of the W43 complex. The RRL emission clearly originates from two distinct velocities ($\sim 40\kms$ and $\sim 100\kms$), with less emission in between. The $\sim 40\kms$ emission is restricted to $\ell > 30.3\degree$, while the $\sim 100\kms$ emission is extended across the entire mapped longitude range. Contours are at 0.25, 0.5, 1, 2, and 4\,K. The green crosses indicate the location of \hii\ regions; the size of each cross is proportional to the RRL intensity of the region. Middle: same, for \hii\ region emission only. Bottom: same, for DIG-only map. \label{fig:lv_w43}}
\end{figure*}

The hydrogen RRL emission at $\sim 40\kms$ and $\sim 100\kms$ is clearly visible in the GDIGS data. We show a total-power spectrum of the mapped area in Figure~\ref{fig:spectrum_w43}. In the top panel of Figure~\ref{fig:lv_w43} we show the longitude--velocity ($\ell$,\,$v$)-diagram of the W43 complex, restricted to $|b| < 0.5$\degree. The $100\kms$ component is distributed across the entire longitude range, while the $40\kms$ emission is mostly restricted to $\ell > 30.3$\degree. The two components are separated in velocity space which may suggest that they arise from two separate gas clouds. It is surprising that there is essentially no $40\kms$ emission near G29 ($\ell \sim 29.9$\degree). Some weak RRL emission is present between $30 - 70\kms$ at $\ell < 29.6$\degree, which is likely caused by radiation escaping from two nearby \hii\ regions outside the mapped area: G029.165--0.035 and G029.110--0.023. The \emph{WISE} catalog lists the coordinates of the \hii\ regions as ($\ell$,\,$b$)=($29.165\degree$,\,$-0.035\degree$) for G029.165--0.035 and ($\ell$,\,$b$)=($29.110\degree$,\,$-0.023\degree$) for G029.110--0.023, and their hydrogen RRL velocities as 61.7 and 52.4\kms, respectively.

\begin{figure*}
\centering
\begin{tabular}{cc}
\includegraphics[width=.5\textwidth]{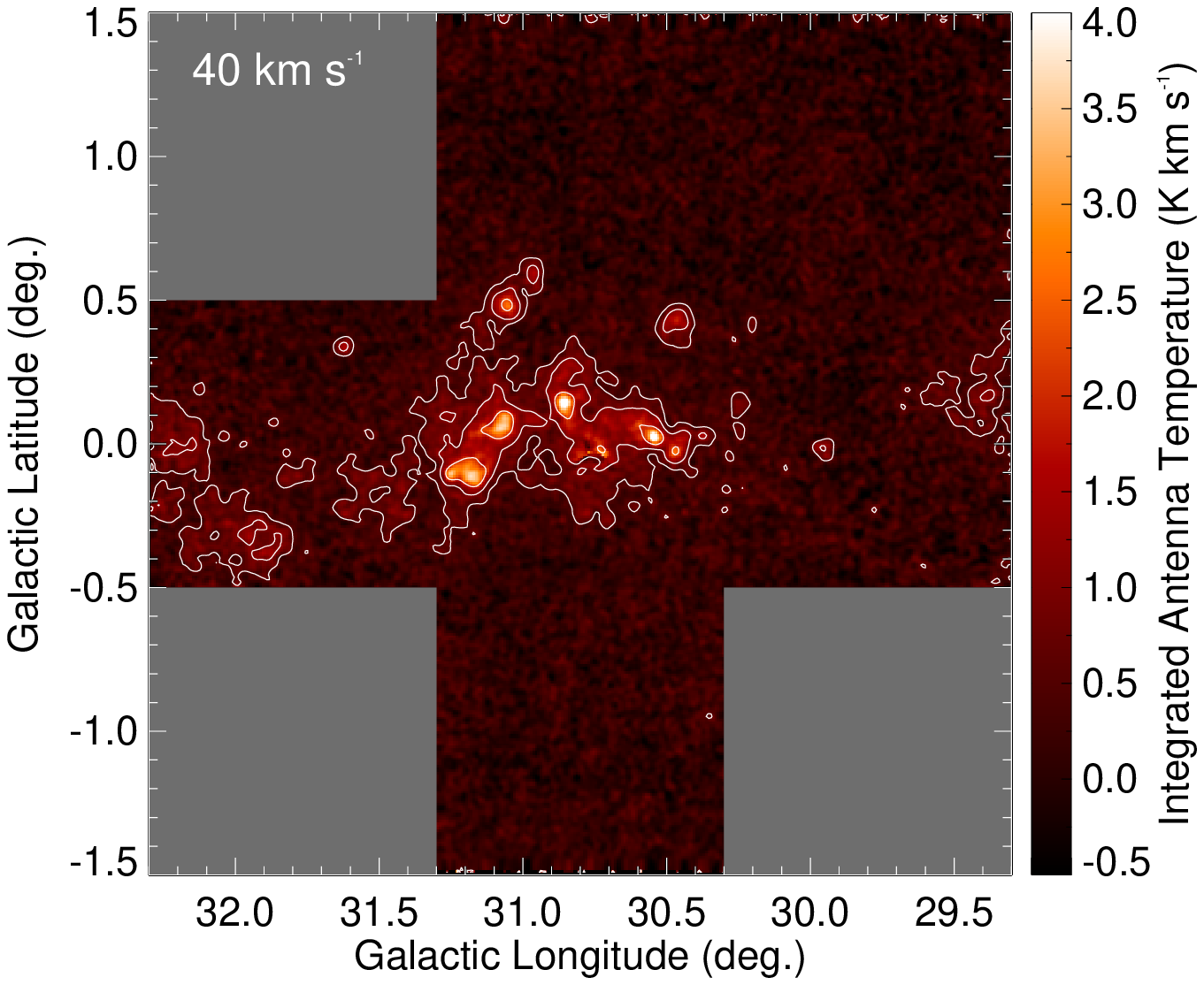}
\includegraphics[width=.5\textwidth]{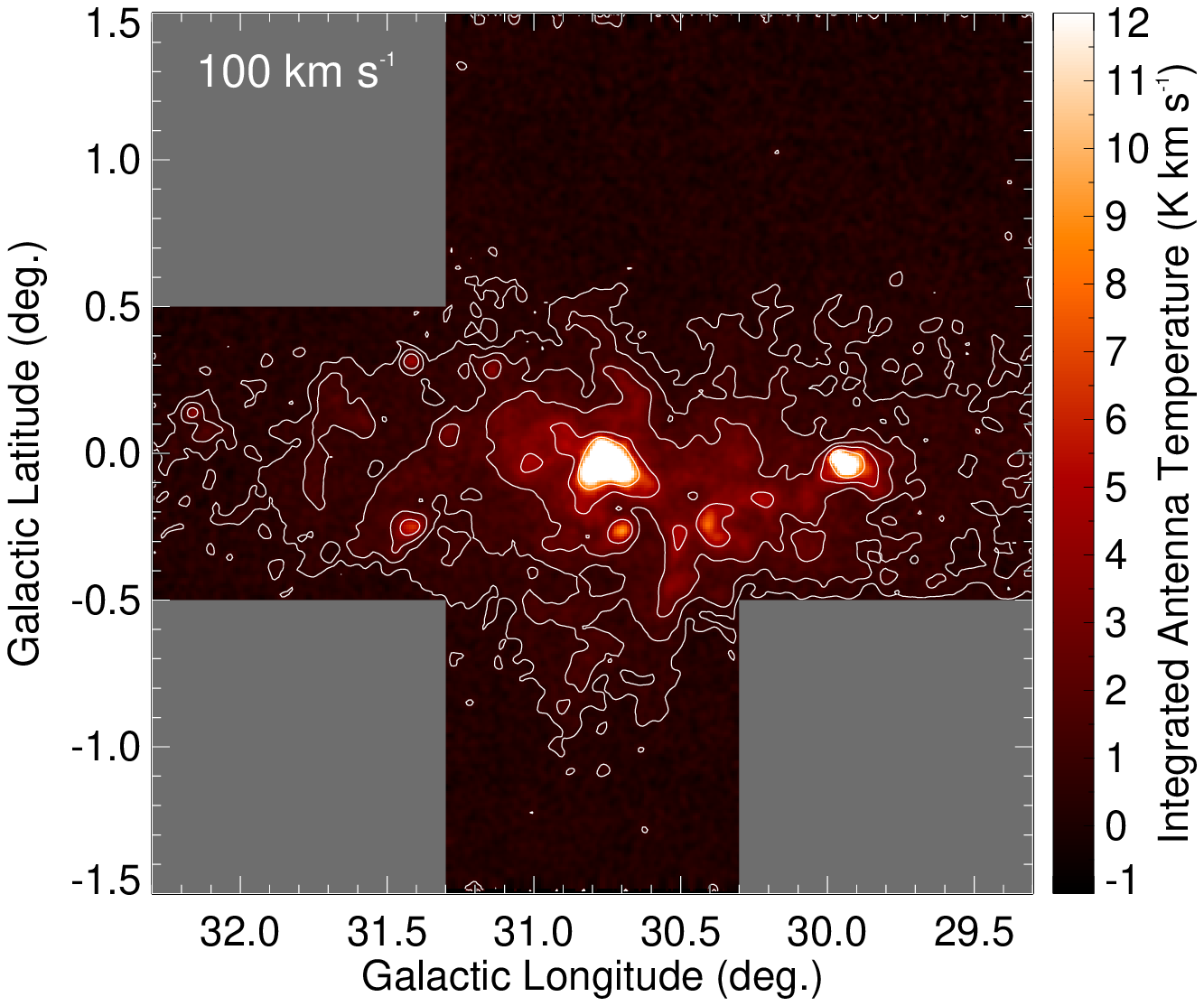}\vspace{-14pt}
\end{tabular}
\caption{Left: Moment 0 map of the W43 complex for the 40\kms velocity component (integrated from 10--65\kms). Contours are at 0.5, 1, 2, 4, and 8\,K\kms. The faint emission at $\ell < 29.6\degree$ is likely due to two nearby \hii\ region outside the mapped area. Right: Same, but for the 100\kms velocity component (integrated from 65--130\kms). \label{fig:mom0_w43_v}}
\end{figure*}

The DIG emission appears to follow the same trends as the overall ionized gas emission. We show the DIG-only ($\ell$,\,$v$)-diagram of the W43 complex in the bottom panel of Figure~\ref{fig:lv_w43}.  We find that in the DIG-only datacube, the $40\kms$ and $100\kms$ components appear more strongly separated in velocity than for the \hii\ region ($\ell$,\,$v$)-diagram shown in the middle panel of Figure~\ref{fig:lv_w43}. This does not necessarily imply, however, that much of the RRL emission between $40$ and $100\kms$ must be due to discrete \hii\ regions. Due to their Gaussian velocity profile with typical RRL widths $\sim 25$\kms, emission from bright \hii\ regions is detectable over a wide velocity range. By removing the \hii\ region component, we also subtract fainter emission at velocities distant from the \hii\ region velocities. As a result, the lack of RRL emission in the $40-100\kms$ range in the DIG-only data may simply highlight the removal of discrete \hii\ regions.

The integrated emission of the DIG-only data associated with the $40\kms$ component (defined as within the 10--65\kms range) is 787\,K\kms, while the integrated $100\kms$ emission (within the 65--130\kms range) is 1{,}994\,K\kms. For the original data including discrete \hii\ regions, the numbers are 1{,}177\,K\kms ($40\kms$ component) and 3{,}933\,K\kms ($100\kms$ component), respectively. The fraction of DIG emission to total emission is larger for the $40\kms$ component (67\%) than the $100\kms$ component (51\%). There is no statistically significant difference between the scale heights of the DIG-only 40 and 100\kms velocity components ($30 \pm 15\arcmin$ and $29 \pm 6\arcmin$, respectively).

\subsection{The origin of the 40\,km\,s$^{-1}$ velocity component}
While the 100\kms DIG component is likely associated with W43 at a distance of $\sim 6$\,kpc, the origin of the 40\kms component is less clear. \citet{Luisi2017} argue that the most likely kinematic distance for the 40 component is  $\sim 12$\,kpc. Based on their pointed RRL data, however, they cannot rule out that the observed velocities near W43 are caused by interacting gas clouds at a single distance. Assuming a combination of unperturbed, purely circular gas motion around the Galactic center and gas streaming motions along the bar, \citet{Luisi2017} derive an expected perturbed velocity component of $57 \pm 4$\kms for the W43 complex, which is near the observed 40\kms emission. An ionized gas component centered near 40\kms is also found in H$\alpha$ emission from the WHAM survey. Based on its vertical morphology, \citet{Krishnarao2017} associate this component with the nearby Sagittarius-Carina spiral arm at a distance of $\sim 2$\,kpc.

We show moment 0 maps of the 40 and 100\kms components in Figure~\ref{fig:mom0_w43_v}. The strongest 40\kms emission is found as an arc-like feature surrounding W43 toward the Galactic north, extending over a range of $\sim 1$\degree in longitude. Fainter 40\kms emission, possibly associated with nearby \hii\ regions, is visible at $\ell \sim 32$\degree. The 40\kms emission component found at $\ell < 29.6$\degree is likely caused by \hii\ regions outside the mapped area (see above).

\begin{figure*}
\centering
\begin{tabular}{cc}
\includegraphics[width=.5\textwidth]{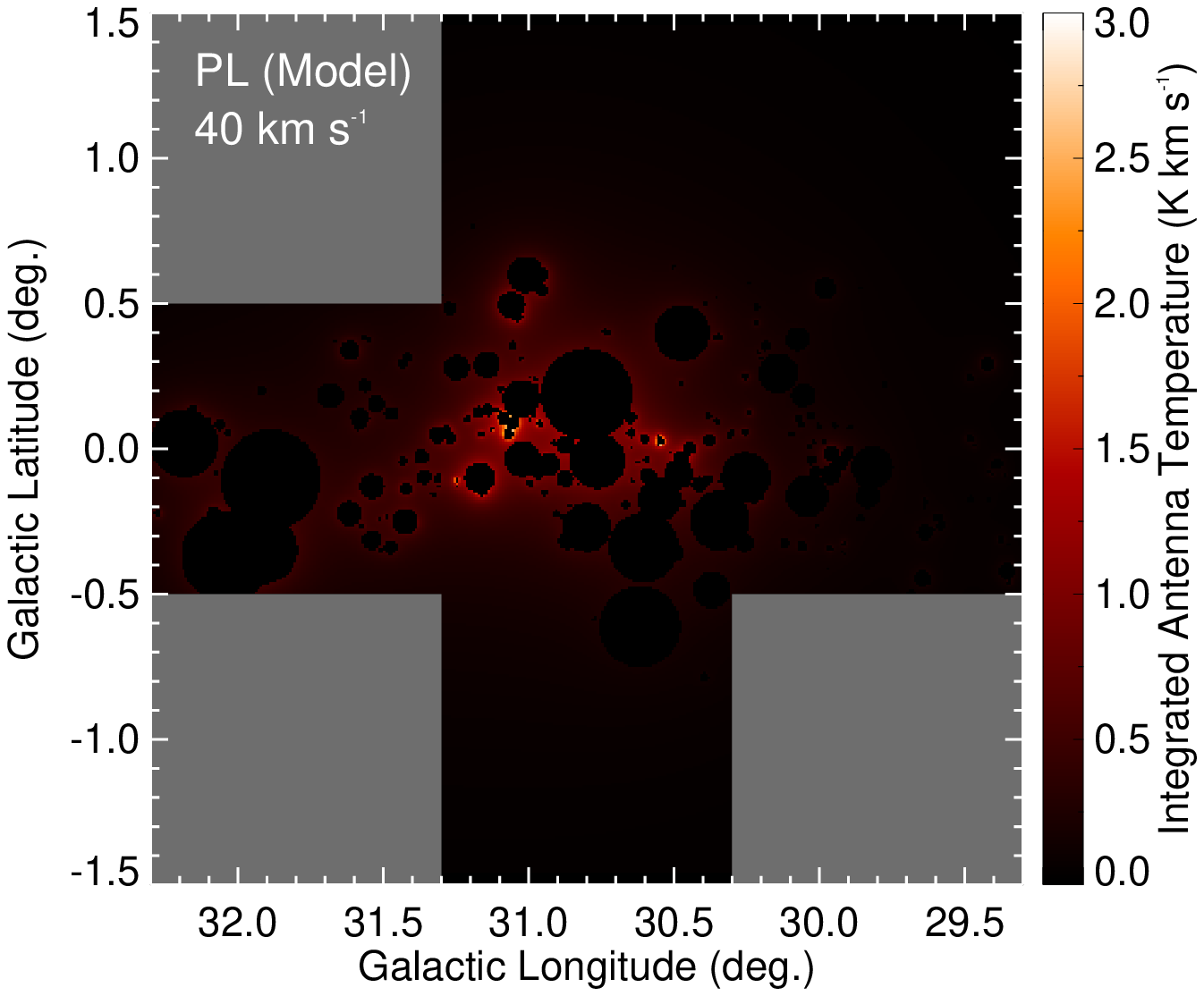} & 
\includegraphics[width=.5\textwidth]{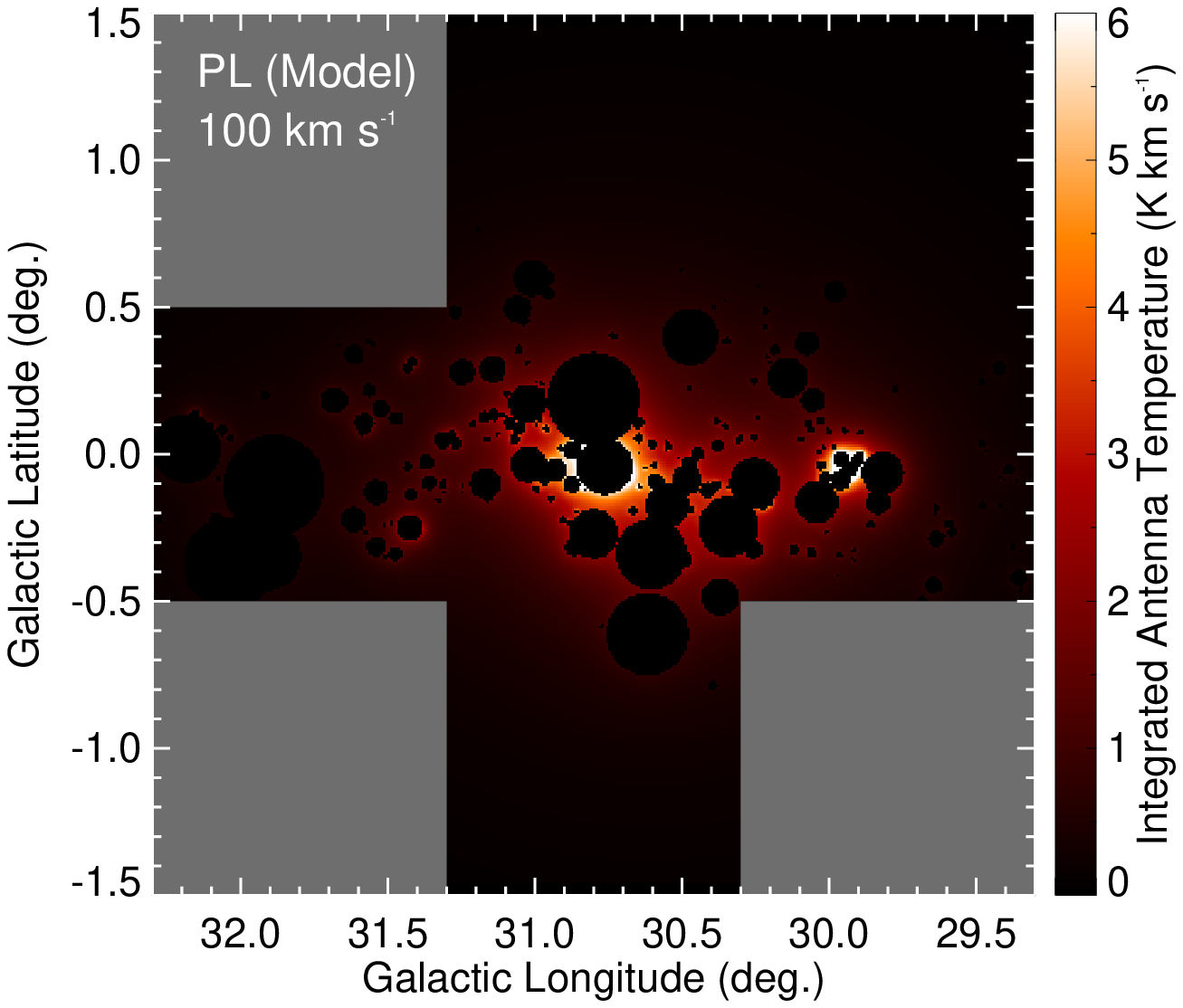}\vspace{-12pt}\\
\includegraphics[width=.5\textwidth]{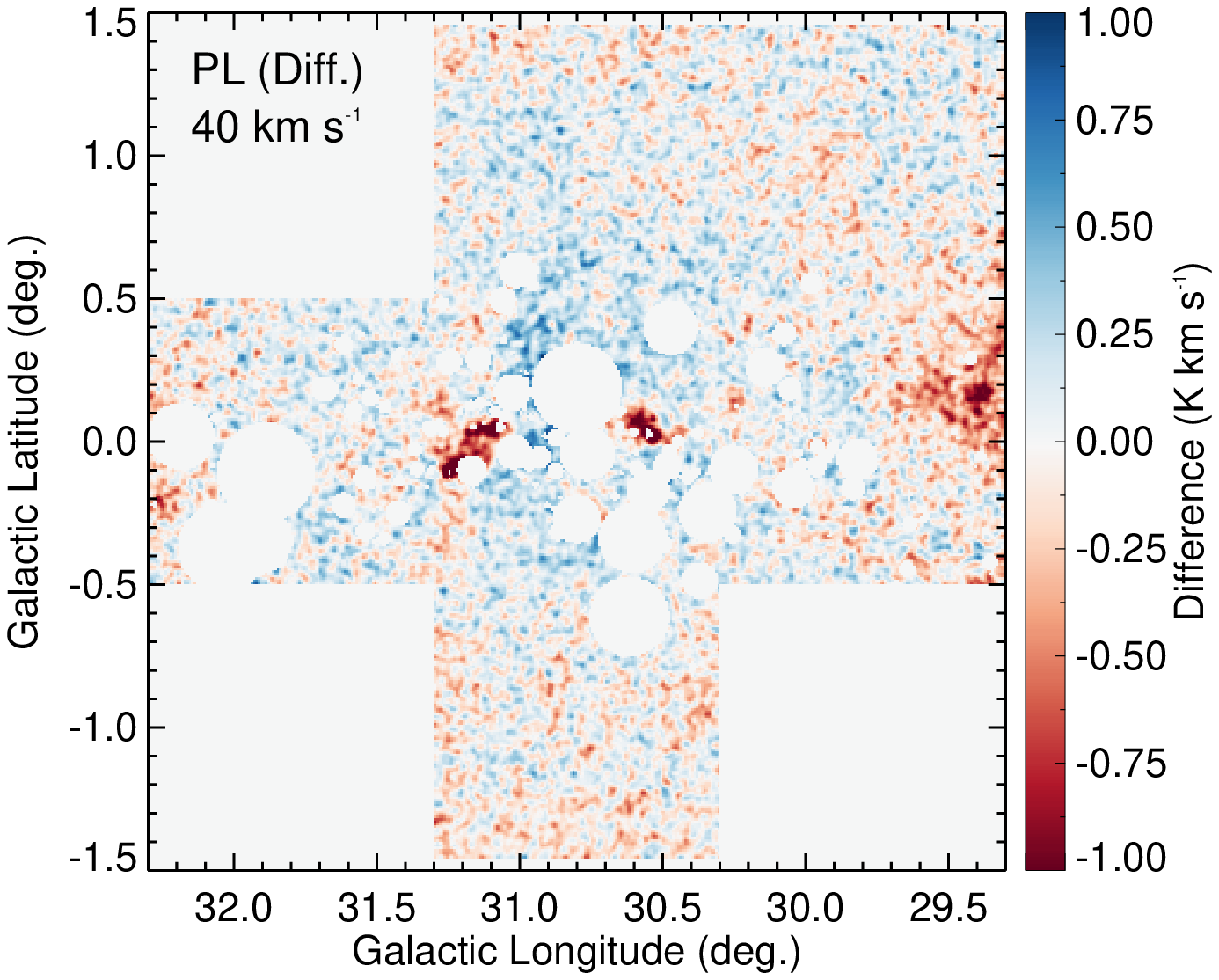} &
\includegraphics[width=.5\textwidth]{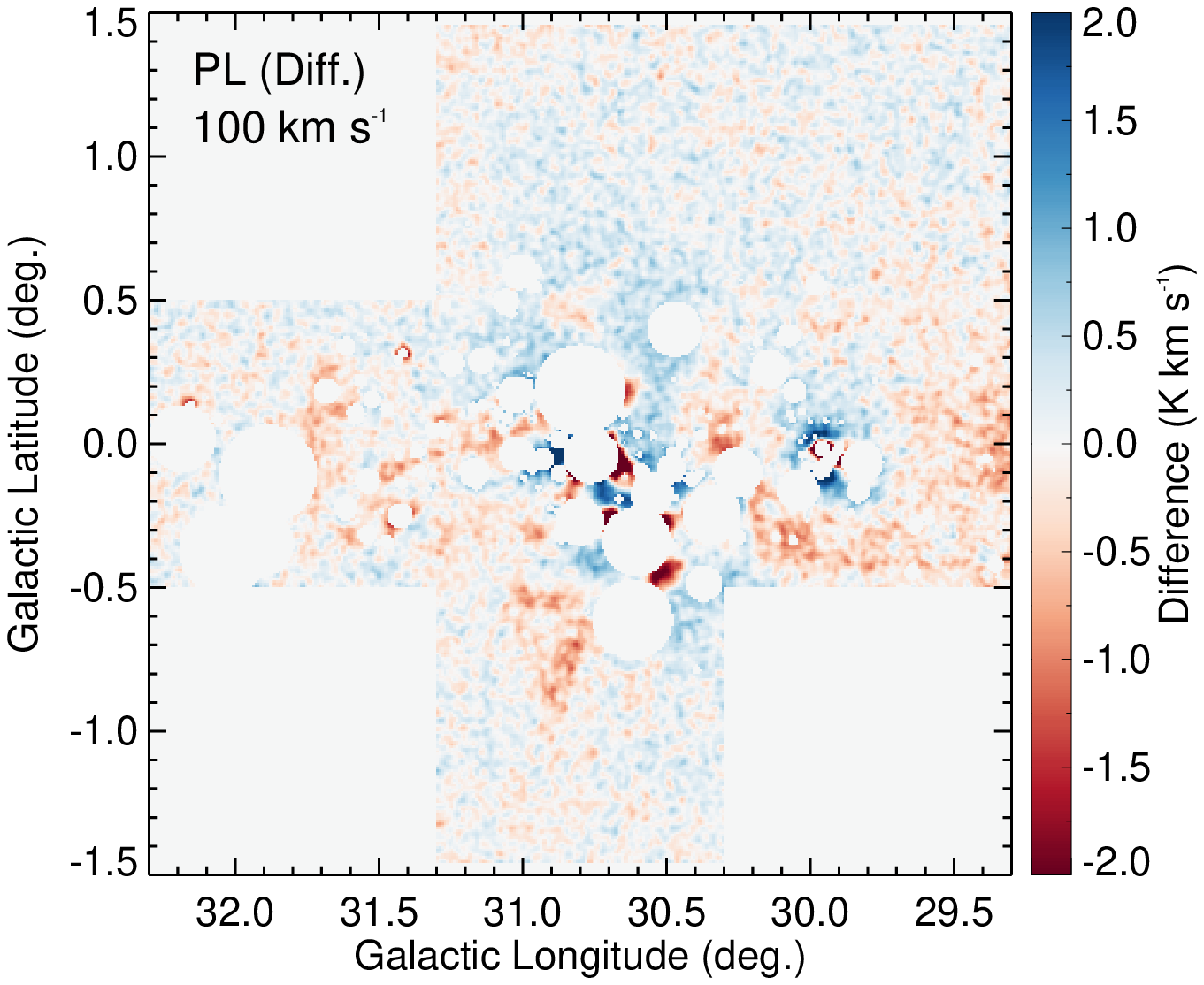}\vspace{-12pt}
\end{tabular}
\caption{Top left: model of diffuse hydrogen RRL emission, for the 40\kms velocity component only, assuming a power-law decrease in RRL emission with distance from each \hii\ region. Bottom left: difference map between the GDIGS moment 0 data and the model. Larger values indicate stronger emission in the model. Top right: same as top left, but for the 100\kms component. The model quality is higher for the 100\kms emission, indicating that the 40\kms DIG may partially maintain its ionization  from \hii\ regions at larger velocities such as W43. This may indicate a small physical separation between the 40\kms DIG and W43. Bottom right: difference map between the GDIGS moment 0 data and the power-law model. \label{fig:model_v}}
\end{figure*}

The spatial confinement of the strong 40\kms emission at the end of the Galactic bar is suggestive of a non-circular origin due to streaming motions around the bar. This would place the observed 40\kms DIG at relatively small physical distances ($\sim 100$\,pc) from the W43 complex. As a result, we would expect W43 and nearby \hii\ regions at similar velocities ($\sim 90$\kms) to contribute to the ionization of the diffuse 40\kms component.

We test this hypothesis by running our model of expected DIG emission (\S \ref{sec:model}) separately for the 40 and 100\kms velocity components. If much of the 40\kms DIG maintains its ionization from sources at larger velocities, we would expect a poor correlation between the model output and the GDIGS data when only considering the DIG and \hii\ regions near 40\kms (within the 10--65\kms range). We show our model output and the difference maps for the two velocity components in Figure~\ref{fig:model_v} (power-law model only) and give our best-fit model parameters in Table~\ref{tab:model} (power-law and exponential model). We find that the correlation between the 100\kms data and the model is indeed much better than that of the 40\kms component (Pearson's $r$ of 0.905 vs.~0.683). We note that the DIG emission found at $\ell < 29.6$\degree may negatively impact our model quality for the 40\kms component. When we blank all pixels with increased DIG emission at $\ell < 29.6$\degree and re-run the model, the correlation is only marginally improved ($r = 0.710$). This suggests that the DIG near 40\kms maintains a significant portion of its ionization from \hii\ regions outside the 10--65\kms range, possibly from W43 itself.

If a significant fraction of the DIG near 40\kms maintains its ionization from W43 and other physically nearby \hii\ regions, it would place the 40\kms emission at a distance of $\sim 6$\,kpc as well. The absence of emission between the two velocity components in $\ell$,\,$v$-space is unexpected, however. The lack of RRL detection between 55 and 75\kms may be explained by a clumpy nature of the 40\kms DIG, connected to the 100\kms component by low-density plasma below our detection threshold.

\section{Ionic Abundances}\label{sec:abundances}
As radiation escapes from an \hii\ region and into the surrounding ISM, the spectrum is affected by absorption and emission processes. Photoionization simulations suggest that helium-ionizing photons become suppressed as the radiation propagates through the \hii\ region while the hydrogen-ionizing continuum is hardened \citep{Wood2004}. The radiation hardening may be due to the much larger ionization cross-section of helium compared to hydrogen, resulting in the absorption of a large fraction of helium-ionizing photons within the ionization front of the \hii\ region \citep{Osterbrock1989}. For O-stars with effective temperatures $< 35{,}000$\,K, \citet{Weber2019} suggest that nearly the entire helium-ionizing radiation is absorbed within the \hii\ region.

The radiation field hardness can be constrained by the $y^+ = \heh$ ionic abundance ratio using RRL observations. Helium is ionized by more energetic radiation ($h \nu \geq 24.6$\,eV) than hydrogen ($h \nu \geq 13.6$\,eV), and therefore a larger value of $y^+$ indicates a harder radiation field. Here,
\begin{equation} y^+ = \frac{T_{\rm L}(\textnormal{He}^+)\Delta V (\textnormal{He}^+)}{T_{\rm L}(\textnormal{H}^+)\Delta V (\textnormal{H}^+)},\label{eq:heh} \end{equation}
where $T_{\rm L}(\textnormal{He}^+)$ and $T_{\rm L}(\textnormal{H}^+)$ are the RRL intensities of helium and hydrogen, respectively, and $\Delta V (\textnormal{He}^+)$ and $\Delta V (\textnormal{H}^+)$ are their FWHM line widths \citep{Peimbert1992}. In a previous study we found that for six out of eight observed \hii\ regions, $y^+$ decreases with distance from the region, suggesting that helium-ionizing photons are indeed suppressed \citep{Luisi2019}. The two regions that did not follow this trend were S206 and M17, which should have essentially no neutral helium within the \hii\ region due to very hard radiation \citep{Balser2006}. A larger sample size, however, is required to determine whether our results are applicable to the entire Galactic \hii\ region population.

A major challenge in determining $y^+$ using our GDIGS data is distinguishing between hydrogen emission and helium emission. The helium line is shifted by $\sim -122$\kms\ from that of hydrogen and, given a spectrum with sufficient S/N ratio, the helium peak can typically be easily separated from the hydrogen peak for a source with a well-defined velocity. The observed hydrogen RRL emission near W43, however, spans a velocity range from $V_{\rm LSR} \approx 0\kms$ to $V_{\rm LSR} \approx 140\kms$ (see Figure~\ref{fig:lv_w43}). At hydrogen velocities between $\sim 0\kms$ and $\sim 20\kms$ both hydrogen and helium emission may therefore be present. In addition, we cannot rule out hydrogen emission at $V_{\rm LSR} < 0\kms$, possibly originating from the Outer Galaxy. Since we are unable to disentangle the low-velocity hydrogen emission from helium emission, we simply define all RRL emission below a velocity of $10\kms$ as due to helium, and emission $> 10\kms$ as due to hydrogen. We find that the exact choice of the cutoff velocity has only a minor impact on the derived $y^+$ values.

We derive $T_{\rm L}(\textnormal{H}^+)\Delta V (\textnormal{H}^+)$ for our averaged Hn$\alpha$ map by integrating the RRL emission over the entire map spatially, and between 10\kms\ and 132\kms\ in velocity space. We likewise calculate $T_{\rm L}(\textnormal{He}^+)\Delta V (\textnormal{He}^+)$ by integrating the emission over the same helium velocities. We discard data at any other velocity to minimize confusion over whether these data should be classified as hydrogen or helium emission. We also discard data at $\ell < 30.3\degree$, $b > 0.8\degree$ where the baseline is poor. We find $y^+ = 0.078$ for the entire map, a value similiar to that found for the \hii\ region population from the HRDS \citep[$y^+ = 0.068 \pm 0.023$; see][]{Wenger2013}.

\begin{figure}
\centering
\includegraphics[width=0.48\textwidth]{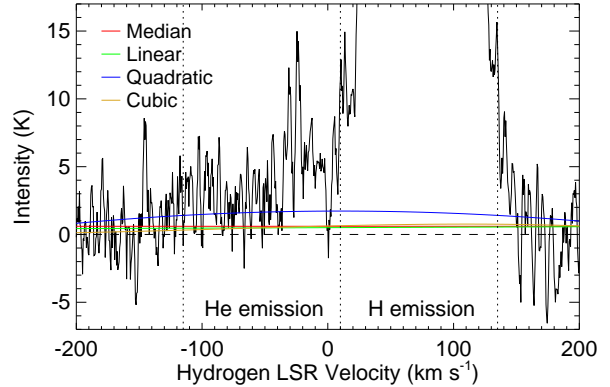}
\caption{RRL spectrum of the mapped area, zoomed in to highlight the effect of fitting baselines of different orders. We fitted the baselines to the same emission-free regions for all pixels and across the entire velocity range of the data ($-350$ to $+350$\kms). The baseline models are shown in color. The dotted vertical lines mark the velocity ranges of expected helium and hydrogen emission. Although the difference between baseline models appears small, the aggregate effect is substantial when deriving $y^+$. Note that the baseline models are not directly fitted to the spectrum shown here, but rather for each individual pixel in the datacube; here we show only the combined effect on the overall spectrum. \label{fig:baselines}}
\end{figure}

\begin{figure*}
\centering
\includegraphics[width=0.89\textwidth]{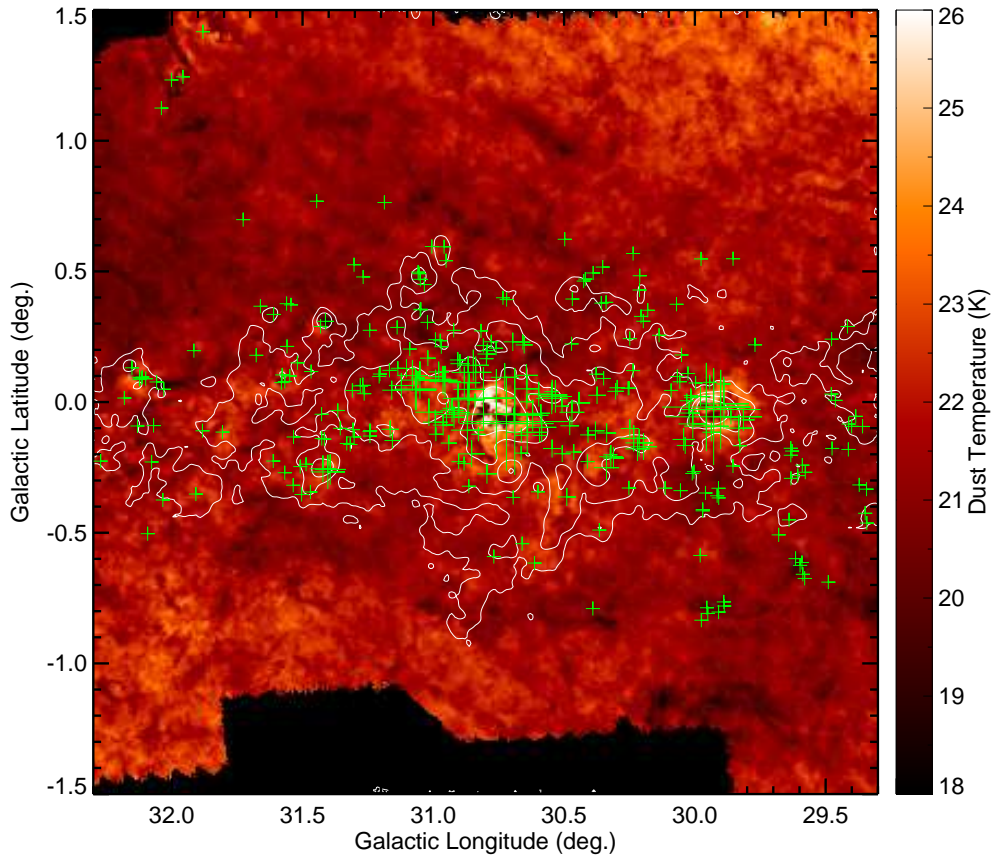}\vspace{-35pt}
\caption{Dust temperature map derived from \emph{IRAS} and \emph{Herschel} data and smoothed to a resolution of 30\arcsec. The contours are of integrated hydrogen RRL intensity and contour levels are at 1, 2, 4, and 8\,K\kms. The green crosses indicate the location of \hii\ regions; the size of each cross is proportional to the RRL intensity of the region. \label{fig:dust_temp}}
\end{figure*}

We repeat the above analysis for the DIG-only datacube (see \S \ref{sec:wimonly}) and find $y^+ = 0.087$ for the DIG, larger than our $y^+$ found for the entire map. This value is in rough agreement with the expected elemental helium abundance in the ISM \citep[$\sim 8.5$\%, see][and references therein]{Balser2006}. We note that the derived value of $y^+$ for the DIG-only data is strongly affected by baseline instability, which we believe is the major uncertainty contribution. In order to quantify the effect of baseline fluctuations, we subtract a 1st to 3rd-order polynomial baseline from our data. The different baseline models change the derived $y^+$ values by up to $\pm 0.028$, although the higher-order baselines likely overfit the data. We show the magnitude of different baseline models in Figure~\ref{fig:baselines}. We therefore conservatively estimate a $\pm 35\%$ uncertainty on all values of $y^+$. Due to the large uncertainty, the difference in $y^+$ between our original averaged Hn$\alpha$ map and the DIG-only data is not statistically significant.

\section{Dust Temperature}\label{sec:dust}
Radiation escaping from \hii\ regions heats the ambient medium. Although dust grains account for only $\sim 1$\% of the mass, they are an important coolant of the ISM since they absorb high-energy photons and re-emit thermally at IR wavelengths. Variations in the dust temperature surrounding \hii\ regions are common. Outside the PDR of the bubble \hii\ region RCW\,120 dust temperature enhancements were found to be correlated with ``holes" in the PDR, along which radiation can escape into the ISM \citep{Anderson2012,Anderson2015}. In a study of the Galactic \hii\ region NGC\,7538 small dust temperature enhancements to the north and east of the region coincide with extended radio emission, suggesting that photons from the \hii\ region leak preferably along these directions \citep{Luisi2016}.

We derive the dust temperature map of the W43 complex shown in Figure~\ref{fig:dust_temp} using 100\,$\mu$m data from the Infrared Astronomical Satellite \cite[\emph{IRAS};][]{Neugebauer1984}, and 250, 350, and 500\,$\mu$m data obtained with the SPIRE instrument on the \emph{Herschel} Space Observatory \citep{Griffin2010} as part of the Hi-GAL Galactic Plane Survey \citep{Molinari2010}. We assume that the dust grain flux is optically thin and follows that of a gray-body,
\begin{equation}
F_{\rm \nu} \propto \kappa _{\rm \nu} B_{\rm \nu}(T_{\rm d}) N_{\rm dust},
\end{equation}
where $F_{\rm \nu}$ is the flux density per beam, $\kappa _{\rm \nu}$ is the dust opacity in cm$^2$\,g$^{-1}$, $B_{\rm \nu}(T_{\rm d})$ is the Planck function for dust temperature $T_{\rm d}$ at frequency $\nu$ (given in Jy\,sr$^{-1}$), and $N_{\rm dust}$ is the dust column density in cm$^{-2}$ \citep{Anderson2012}. We re-grid the IR maps to the resolution of the GDIGS data (30\arcsec) and fit the gray-body model pixel-by-pixel, assuming that the
functional form for the dust opacity is $\kappa _{\rm \nu} = 0.1\,(\nu / 10^{12} \, {\rm Hz}) ^{\beta}$ \citep{Beckwith1990}, with a dust emissivity index of $\beta = 2$. We also fit the column density of gas and dust simultaneously with the dust temperatures, assuming 
\begin{equation}
N_{\rm H} = R \, \frac{F_{\rm \nu}}{2.8 \, m_{\rm H} \kappa _{\rm \nu} B_{\rm \nu}(T_{\rm d}) \Omega},
\end{equation}
where $R=100$ is the gas-to-dust mass ratio, $m_{\rm H}$ is the mass of a hydrogen atom in g (with the factor of 2.8 accounting for heavier elements), and $\Omega$ is the beam size in sr \citep{Anderson2012}.

The average dust temperature in the field of view is $\sim 21$\,K, with no significant large-scale trends (see Figure~\ref{fig:dust_temp}). The two most luminous \hii\ regions on the map, W43 and G29, both have generally higher dust temperatures than average, but also show cold patches. These cold areas correspond to larger column densities, indicating cold dust clumps foreground to the \hii\ regions \citep[see][]{Bally2010}. Apart from W43 and G29, there are a number of \hii\ regions with derived dust temperatures lower than average and large column densities in combination with above-average $T_{\rm d}$ values from their surroundings (e.g., the \hii\ region complex G032.110$+$00.090 at $\ell \approx 32.1$\degree, $b \approx 0.1$\degree and G031.401--00.259 at $\ell \approx 31.4$\degree, $b \approx -0.3$\degree). This suggests that these regions are still embedded in molecular clouds, but that a fraction of their radiation escapes into the ISM and heats the diffuse gas. Many \hii\ regions in Figure~\ref{fig:dust_temp} have the same dust temperature as their surroundings. Most of these regions have integrated RRL intensities $< 1$\,K\kms, which may indicate that their luminosities are insufficient to have a measurable effect on the derived dust temperature. 

\begin{figure}
\centering
\includegraphics[width=0.49\textwidth]{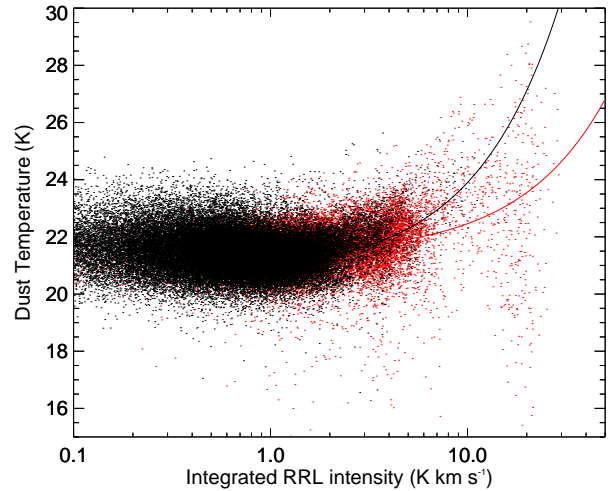}
\caption{Pixel-by-pixel correlation between integrated hydrogen RRL intensity and dust temperature. The red dots mark emission coincident with discrete \hii\ regions, while the black dots mark diffuse gas emission. The red and black lines are linear fits of the form $T_{\rm d} = a + b \, I_{\rm RRL}$ to the data, where $T_{\rm d}$ is the dust temperature in K and $I_{\rm RRL}$ is the integrated RRL intensity in K\kms. We only fit data points with $I_{\rm RRL} > 1$\,K\kms. Fit parameters are $a = 21.37 \pm 0.01$, $b = 0.109 \pm 0.003$ for the correlation between \hii\ region RRL emission and $T_{\rm d}$, and $a = 20.68 \pm 0.02$, $b = 0.322 \pm 0.009$ for the diffuse gas. Note that the GBT beam size is larger than the pixel size and therefore the intensities of pixels within one beam area are not truly independent of each other.\label{fig:dust_corr}}
\end{figure}

We show the pixel-by-pixel correlation of integrated RRL intensity and dust temperature in Figure~\ref{fig:dust_corr}. We fit the correlation between RRL emission and $T_{\rm d}$ separately for directions coincident with discrete \hii\ regions and the diffuse ISM, and find that there is a correlation between dust temperature and RRL intensity at integrated intensities greater than $\sim 1$\,K\kms. This may suggest that the same radiation field that heats the dust also maintains the ionization of the DIG.  The slope of the correlation is steeper for diffuse locations. The low dust temperatures ($T_{\rm d} < 20$\,K) at RRL intensities $>$10\,K\kms\ are mostly due to the cold dust clumps assumed to be foreground to W43 and G29. Since large differences in dust heating mechanisms within \hii\ regions and the diffuse gas are unlikely, these cold foreground clumps may be the principal cause for the distinct slopes, since they artificially introduce low dust temperatures (from the foreground clumps) associated with large RRL intensities (from the background star-forming regions). There is no statistically significant correlation between RRL intensity and dust temperature at lower RRL intensities.

\section{Summary}\label{sec:conclusions}
Using data from the GDIGS survey, an ongoing RRL survey at C-band (4--8\,GHz) with the GBT, we perform a high-resolution spectroscopic study of the diffuse ionized gas in the Galactic plane. Here we present our GDIGS data of W43, one of the most active star formation regions in the inner Galaxy. There are roughly 300 \hii\ regions and \hii\ region candidates within $\sim 1.5$\degree of W43, making the area an ideal target to study the impact of \hii\ regions on the surrounding ISM. 

Using our fully-sampled RRL observations from the GDIGS survey and data from the \emph{WISE} Catalog of Galactic \hii\ Regions \citep{Anderson2014}, we generate DIG-only maps to analyze the properties of the diffuse gas unaffected by emission from discrete \hii\ regions. We find that the emission from the DIG is strongest near W43 and G29, but there is also significant diffuse RRL emission at large distances from the Galactic plane far from any known active star-forming regions. One such emission region is $\sim 1\degree$ to the Galactic south of W43, which may be associated with the ``worm-ionized medium" postulated by \citet{Heiles1996a}. This structure may indicate that there exist low-density pathways perpendicular to the Galactic plane, allowing Lyman continuum photons to travel the distances necessary to account for the observed DIG scale height of $\sim 1$\,kpc.

The intensity of the DIG emission can be estimated by a simple empirical model based on the locations, angular sizes, and RRL intensities of the \hii\ regions in the field of view. As a first-order approximation, the model assumes that all DIG emission is due to photons leaking from \hii\ regions. We find that the correlation between model emission and our GDIGS RRL data is improved by assuming a power-law decrease in hydrogen RRL emission beyond the \hii\ region PDRs rather than an exponential decrease.

Most of the RRL emission near W43 and G29 is found at two distinct velocities: 40 and 100\kms. The distance determination for the ionized gas responsible for these two velocity components is hampered by their location in the inner Galaxy and possibly by strong non-circular motions near the end of the bar where W43 is located. Parallax measurements place W43 at a distance of $\sim 5.5$\,kpc \citep{Zhang2014}. Since the 100\kms gas component is spatially associated with W43 we argue that it is at roughly the same distance. The correlation between the observed and the modeled 40\kms RRL emission is poor, suggesting that the 40\kms component may partially maintain its ionization from \hii\ regions at larger velocities. We therefore hypothesize that the two velocity components are interacting, placing the 40\kms emission at a distance of $\sim 6$\,kpc as well.

We use the helium-to-hydrogen ionic abundance ratio, $y^+$, as a tracer for the hardness of the radiation field, however, we do not find a statistically significant difference in $y^+$ between our original averaged Hn$\alpha$ map and the DIG-only data. The correlation between dust temperature and RRL intensity may suggest that the same radiation field that heats the dust also maintains the ionization of the DIG.

\acknowledgments
We thank the anonymous referee for useful comments and suggestions, which improved the quality of this manuscript. \gboblurb\ \nraoblurb\ \emph{Herschel} is an ESA space observatory with science instruments provided by European-led Principal Investigator consortia and with important participation from NASA. We thank West Virginia University for its financial support of GBT operations, which enabled some of the observations for this project. This work is supported by NSF grants AST1812639 to LDA and AST1714688 to TMB. BL thanks West Virginia University for hosting him as a postdoctoral researcher during completion of this work. BL was partially supported by the National Natural Science Foundation of China (Grant \#11503036).\\

\textit{Facility:} Green Bank Telescope.
\textit{Software:} GBTIDL \citep{Marganian2006}, gbtgridder (https://github.com/ \\
GreenBankObservatory/gbtgridder).

\bibliographystyle{aasjournal}
\bibliography{HII}

\end{document}